\documentclass[
aps
,prx
,twocolumn,10pt
,superscriptaddress
,longbibliography
,amsmath
,amssymb
,amsfonts
,floatfix
]{revtex4-2}
\usepackage{float}
\usepackage{commath}
\usepackage{graphicx}
\usepackage{verbatim}
\usepackage{color}
\usepackage[dvipsnames,svgnames,x11names,hyperref]{xcolor}

\usepackage[normalem]{ulem}
\usepackage{times}
\usepackage[sf,tight]{subfigure}

\usepackage{siunitx}
\DeclareSIUnit\gauss{G}

\definecolor{linkcolor}{RGB}{6,69,173} 
\definecolor{diffcolor}{RGB}{175,31,36} 

\usepackage[colorlinks=true,
            linkcolor=linkcolor,
            urlcolor=linkcolor,
            citecolor=linkcolor,
            unicode,
            pdfencoding=auto]{hyperref}

\usepackage[
]{physpack}

\newcommand{\ab}[1]{{\color{black} #1}}

\newcommand{\COMMENT}[1]{}

\newcommand{\cezr}{Ce$_2$Zr$_2$O$_7$}
\newcommand{\cehf}{Ce$_2$Hf$_2$O$_{7}$}
\newcommand{\lahf}{La$_2$Hf$_2$O$_{7}$}

\begin{document}

\title{Thermodynamics of the dipole-octupole pyrochlore magnet \cehf~in applied magnetic fields}

\author{Anish Bhardwaj}
\email{anis.bhardwaj@gmail.com}
\affiliation{Department of Physics, Florida State University, Tallahassee, Florida 32306, USA}
\affiliation{National High Magnetic Field Laboratory, Tallahassee, Florida 32310, USA}
\affiliation{Department of Physics, St. Bonaventure University, New York 14778, USA}
\author{Victor Por\'{e}e}
\email[]{victor.poree@psi.ch}
\affiliation{Laboratory for Neutron Scattering and Imaging, PSI Center for Neutron and Muon Sciences, Paul Scherrer Institut, 5232 Villigen PSI, Switzerland}
\affiliation{Synchrotron SOLEIL, L'Orme des Merisiers, Saint-Aubin, BP 48, F-91192 Gif-sur-Yvette, France}
\author{Han Yan}
\affiliation{Department of Physics \& Astronomy, Rice University, Houston, TX 77005, USA}
\affiliation{Smalley-Curl Institute, Rice University, Houston, TX 77005, USA}
\author{Nicolas~Gauthier}
\affiliation{Institut Quantique, D\'epartement de physique, and RQMP, Universit\'e de Sherbrooke, Sherbrooke, Qu\'ebec J1K 2R1, Canada}
\author{Elsa Lhotel}
\affiliation{Institut N\'eel, CNRS, Universit\'e Grenoble Alpes, 38042 Grenoble, France}
\author{Sylvain~Petit}
\affiliation{LLB, CEA, CNRS, Universit\'{e} Paris-Saclay, CEA Saclay, 91191 Gif-sur-Yvette, France}
\author{Jeffrey A. Quilliam}
\affiliation{Institut Quantique, D\'epartement de physique, and RQMP, Universit\'e de Sherbrooke, Sherbrooke, Qu\'ebec J1K 2R1, Canada}
\author{Andriy H. Nevidomskyy}
\affiliation{Department of Physics \& Astronomy, Rice University, Houston, TX 77005, USA}
\affiliation{Rice Center for Quantum Materials and Advanced Materials Institute, Rice University, Houston, Texas 77005, USA}
\author{Romain Sibille}
\email[]{romain.sibille@psi.ch}
\affiliation{Laboratory for Neutron Scattering and Imaging, PSI Center for Neutron and Muon Sciences, Paul Scherrer Institut, 5232 Villigen PSI, Switzerland}
\author{Hitesh J. Changlani}
\email{hchanglani@fsu.edu}
\affiliation{Department of Physics, Florida State University, Tallahassee, Florida 32306, USA}
\affiliation{National High Magnetic Field Laboratory, Tallahassee, Florida 32310, USA}

\begin{abstract} 
{
The recently discovered dipole-octupole pyrochlore magnet Ce$_2$Hf$_2$O$_7$ is a promising three-dimensional quantum spin liquid candidate which shows no signs of ordering at low temperature. 
The low energy effective pseudospin-1/2 description in a magnetic field is characterized by the XYZ Hamiltonian and a Zeeman term where the dipolar local $z$-component of the pseudospin couples to the local $z$-component of the applied magnetic field, while the local $x$- and $y$-components of the pseudospin remain decoupled as a consequence of their octupolar character.
Using effective Hamiltonian parameters determined in 
V. Poree \textit{et al.}, arXiv:2305.08261 (2023),
remarkable experimental features can be reproduced, as for instance the specific heat and magnetization data as well as the continuum of states seen in neutron scattering. Here we investigate the thermodynamic response to magnetic fields applied along the global [110]
direction using specific heat measurements and fits using numerical methods, and solve the corresponding magnetic structure using neutron diffraction. Specific heat data in moderate fields are reproduced well, 
however, at high fields the agreement is not satisfactory. We especially observe a two-step release of entropy, a finding that demands a review of both theory and experiment. 
We address it 
within the framework of three possible scenarios, including an analysis of the crystal field Hamiltonian not restricted to the two-dimensional single-ion doublet subspace. We conclusively rule out two of these scenarios and find qualitative agreement with a simple model of field misalignment with respect to the crystalline direction. 
We discuss the implications of our findings for [111] applied fields and for future experiments on Ce$_2$Hf$_2$O$_7$ and its sister compounds. 
}
\end{abstract}

\maketitle

\section{Introduction} 
While initially viewed as a completely theoretical enterprise, the study of quantum spin liquids (QSLs)   \cite{SUBRAMANIAN1983,Hermele:2004gg,Balents2010,lacroix2011introduction,Gingras:2014ip,GardnerRevModPhys,RauGingrasAnnuRevCMP,HallasXYAnnuRevCMP,Yan2017PhysRevB} has seen considerable excitement and research activity in condensed matter physics. This exotic phase of matter is characterized by the presence of fractional excitations and general lack of local order parameter, thus requiring a description beyond the Landau-Ginzburg-Wilson paradigm. QSLs are now actively being sought after in real materials thanks to advances in synthesis and crystal growth, and guidance from theory. Some candidate QSL materials include herbertsmithite (kagome structure) \cite{Norman2016RevModPhys,Han2012Nature}, ruthenium trichloride (Kitaev honeycomb material) \cite{ONG,rucl31,rucl32,Takagi-quantized2022,kitaev2006,Shuyi2023arxiv}, ${\mathrm{Ca}}_{10}{\mathrm{Cr}}_{7}{\mathrm{O}}_{28}$ (bilayer breathing kagome magnet) \cite{Balz2016,Balz2017PhysRev,PhysRevB.104.024426,Yan2022PhysRevResearch}, and members of the vast family of pyrochlores~\cite{RauGingrasAnnuRevCMP}.

In pyrochlores, the magnetic ions form a lattice of tetrahedra connected by their summits, and the basic motivation came from spin ice physics, originally discovered in Ho$_2$Ti$_2$O$_7$ and Dy$_2$Ti$_2$O$_7$ \cite{Fennell2009,Morris2009,Ramirez_1999,Anand2022}. In these materials, a strong crystal field forces dipole magnetic moments to align along or against the local $z$ axis (local $[111]$ axis) directed from the site to the center of the tetrahedron. Such systems do not show any long range order down to mK temperatures, instead their properties are governed by magnetic configurations formed by an organizing principle, the `2-in 2-out ice rule', emerging from effective ferromagnetic interactions. This ice rule constraint manifests itself in various ways, notably the observation of Pauling's residual entropy~\cite{Ramirez_1999} and the existence of a classical Coulomb liquid with characteristic `pinch points'~\cite{henley_coulomb} accessible in neutron scattering. 

One approach to find QSLs has been to look for pyrochlore magnets with relaxed constraint on the Ising anisotropy of the magnetic moments. This approach, however, has proved difficult. For example, the case of Yb$_2$Ti$_2$O$_7$~\cite{YTO.Ross,YTO.applegate} was anticipated to be a QSL but was found to have a ferromagnetic ground state~\cite{Chang2012,YTO.Scheie,YTO.Scheie_PRL,Rau2019,Saubert2020}. Alternatively, a promising route for QSL physics was found with materials based on the non-Kramers trivalent Pr$^{3+}$ ion having Ising-like dipole magnetic moments~\cite{zhou08,Kimura:2013gj,Princep2013,petit16-PRB94,PHO_2016,wen17-PRL118,anand16-PRB94,savary17-PRL118,PHO_NatPhy,PhysRevLett.121.037203,Tang2023}, and where interactions between electric quadrupoles~\cite{onoda10-PRL105} and residual non-magnetic disorder~\cite{savary17-PRL118,PhysRevLett.121.037203} can both be source of quantum fluctuations to stabilize a QSL.

More recently, the case of trivalent rare-earth ions that can stabilize a `dipole-octupole' ground state doublet, such as Nd or Ce, has been actively investigated and also appears very promising~\cite{CZO.Gao, CZO.Gaulin, natphys.sibille, Petit16, Leger_2021}. As in most rare-earth pyrochlores, since the crystal-electric field (CEF) ground state doublet is well separated from the rest of the spectrum by a gap of few tens of meV, microscopic descriptions use an effective spin 1/2 degree of freedom. The special feature of these Kramers dipole-octupole doublets lies in the fact that the $z$ component of the pseudo-spin essentially represents the magnetic (dipolar) moment, while the the $x$ and $y$ components are octupolar in nature. This has important consequences on the nature of the low energy effective description. Following previous studies, the theoretical calculations presented in this work are based on a minimal nearest neighbor (nn) XYZ Hamiltonian incorporating all symmetry-allowed effective interactions between these pseudo-spins on the pyrochlore lattice~\cite{huang2014, Patri-octupolarQSI, PMB, ybk_felix_phase, sanders2024do,Bhardwaj2022NatMat}. Accounting for the applied magnetic field when projected onto the ground doublet subspace gives:
\begin{eqnarray}
H_{nn} &=& \sum_{\langle ij\rangle} J_x s_i^x s_j^x + J_y s_i^y s_j^y +  J_z s_i^z s_j^z + J_{xz}(s_i^xs_j^z + s_i^z s_j^x) \nonumber\\
&&-g\mu_B \sum_i (\bfh \cdot \hat{z}_i) (g_x s_i^x + g_z s_i^z).
\label{eq:Hnn}
\end{eqnarray}
The basis is chosen such that $s_i^z$ is defined in the local $[111]$ direction on a given rare-earth site and $g_x=0$. $J_x,J_y,J_z$ and $J_{xz}$ are effective coupling strengths and $\langle ij \rangle$ refers to nearest neighbor pairs of sites. We note that, in the chosen basis with $g_x=0$, the Zeeman term couples only the local $z$ component of the magnetic field to the dipolar $z$ component while the octupolar ($x$ and $y$) components do not contribute to this term.

Depending on the relative strength of $J_x,J_y,J_z$, different situations can emerge. For instance, in Nd$_2$Zr$_2$O$_7$, $J_z$ is dominant and negative, leading to an antiferromagnetic all-in--all-out ground state, with however significant dynamics arising from a non-negligible $J_x$ `transverse' term \cite{Petit16, Benton16, Xu_2020, Leger_2021}. In the Ce$^{3+}$-based pyrochlore materials Ce$_2$Sn$_2$O$_7$, Ce$_2$Zr$_2$O$_7$ and Ce$_2$Hf$_2$O$_7$~\cite{natphys.sibille,prl.sibille,CZO.Gao,CZO.Gaulin,cehfo-poree,Bhardwaj2022NatMat,PoreeArxiv2023,YahneCSO_PRX}, previous work determined a dominant positive interaction between $x$ and/or $y$ pseudo-spin components, hence a new class of spin ices where the ice rule is `octupolar'. 
As a result, the magnetic diffuse scattering in Ce$_2$Sn$_2$O$_7$ does not show the familiar characteristic signature observed in the dipolar spin ice systems. Some  magnetic diffuse scattering, however, arises at high momentum transfers and is observable with neutrons thanks to the weak coupling of octupoles to neutrons ~\cite{natphys.sibille,Bhardwaj2022NatMat}. Furthermore, in all three Ce$^{3+}$-based pyrochlores, a continuum of spectral weight~\cite{CZO.Gao,CZO.Gaulin} -- the hallmark of fractionalized excitations in a QSL, has been observed in neutron scattering in zero applied magnetic field. 

For \cehf, effective exchange parameters were reported previously by us \cite{PoreeArxiv2023} with a dominant $J_y$ interaction:
\begin{equation}
\label{eqn:parameters}
\begin{split}
    &J_x=0.02,\ J_y=0.047, \  J_z=0.013 \\
    & J_{xz}=-0.008,\   g_x=0,\ g_z=2.328
\end{split}
\end{equation}
(all $J$'s in meV, $g$ dimensionless), which locates the material of the present study in the $\pi$-flux U(1) QSL phase \cite{Lee2012PhysRevB,huang2014,GangChen2017PhysRevB, Li_Chen_2017, chen-li-PRR2020,PMB,Patri-octupolarQSI}, as suggested also for Ce$_2$Zr$_2$O$_7$~\cite{CZO.Gaulin,CZO.Gao,Bhardwaj2022NatMat,smith2023}.
The interaction parameters result in a rich phase diagram for cerium pyrochlores, and their tiny energy scales result in a potentially large response to small changes in the material composition. For instance, this is illustrated in the markedly different exchange parameters in Ce$_2$Zr$_2$O$_7$ and Ce$_2$Hf$_2$O$_7$ despite Zr$^{4+}$ and Hf$^{4+}$ sharing very similar ionic radii and electronegativities. The highly tunable nature of these tiny energy scales may also explain the contrasted magnetic neutron scattering response of Ce$_2$Sn$_2$O$_7$ samples prepared at lower temperature~\cite{YahneCSO_PRX} compared to other studies~\cite{prl.sibille,natphys.sibille,prl.sibille,PoreeArxiv2023_CSO}. 
The methodology to deduce reliable sets of exchange parameters from the available experimental data is a central problem in the study of these materials.

In the case of dipolar systems, the application of external magnetic fields has been a valuable resource for building confidence in the microscopic description of such materials~\cite{YTO.Ross,Anand2022}. Large fields spin polarize the system allowing direct measurement of the energy- and momentum-resolved spin waves (for example, see the case of Yb$_2$Ti$_2$O$_7$~\cite{YTO.Ross,YTO.Scheie, YTO.Scheie_PRL, YTO.applegate} and references therein). Using linear spin wave theory, it is then possible to accurately determine the underlying interactions in these dipolar spin ices. The case of dipolar-octupolar magnets is starkly
different~\cite{huang2014,GangChen2017PhysRevB, Li_Chen_2017, chen-li-PRR2020,PMB,Patri-octupolarQSI}: a field-induced magnetic order can exist, but with no spectral weight associated with the octupolar low energy excitations measured with neutrons at low momentum transfers. 
Absence of spin waves has also been noted in Ce$_2$Zr$_2$O$_7$, despite the measurement of a field-induced magnetic order in a finite magnetic field~\cite{smith2023}
These features outline the importance of methods to evaluate exchange interactions of cerium pyrochlores from measured data.

Given this prelude to dipole-octupole pyrochlore magnets, in this work we present neutron diffraction, magnetization and specific heat data on \cehf\ for magnetic fields applied along the crystallographic [110] direction. Strikingly, we infer a two-step release of entropy from the specific heat measurements, a finding that demands a review of the experiment and exploration of multiple theoretical scenarios. Using a combination of quantum many-body calculations and analytical insights, we 
present strong evidence for the importance of a field misalignment away from the [110] axis as the origin of the peculiar evolution of entropy as a function of the applied field strength. We also revisit results for the [111] field direction, previously addressed elsewhere~\cite{PoreeArxiv2023}.

The rest of the paper is organized as follows. In section~\ref{sec:setup} we describe the setup of the pyrochlore lattice, discussing the phases expected for high magnetic fields. In section~\ref{sec:experiments} we discuss important details of our experiments, and the results of measurements for magnetic fields, highlighting the agreement between new data for moderate [110] fields with the prediction from our previously determined model parameters given in Eq.~\eqref{eqn:parameters}. Section~\ref{sec:110_theory} is devoted to our theoretical analysis for [110] fields, while Section~\ref{sec:111} provides, for completeness, a similar analysis for the [111] fields, with simple torque calculations confirming that the effects of a misalignment are far less important in that case. Finally, in Section \ref{sec:conclusion} we conclude and discuss the implication of our work on future experiments. 


\section{Pyrochlore Setup}
\label{sec:setup}
\begin{figure}[htpb]
\includegraphics[width=\linewidth]{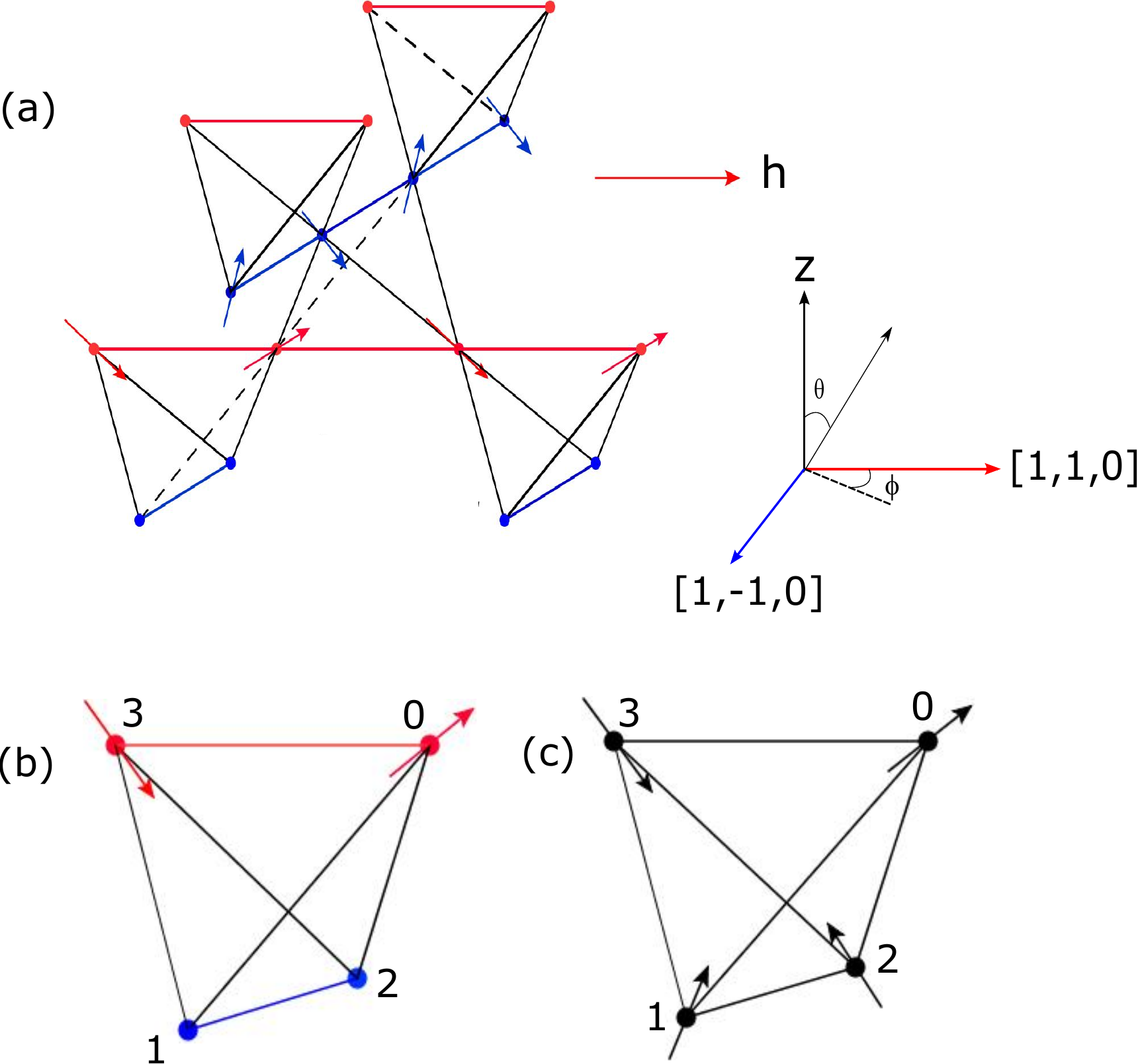}
\caption{(a) Schematic of a pyrochlore lattice in the presence of an applied [110] magnetic field (indicated by h). The red (blue) solid circles represent the $\alpha (\beta)$ sites. Each $\alpha/\beta$ chain is represented by a red/blue line directed along the [110]/[1-10] direction. A general field direction can be uniquely specified by the angle ($\phi$) it makes with the [110] direction and the angle ($\theta$) it makes with the global z-axis. (b) and (c) depict the ground state configuration of the pseudospins in the presence of very high field directed along the [110] and [111] directions respectively. There is no net moment on the $\beta$ sites in the former case, while the latter shows the typical `3-in--1-out' (`1-in--3-out') state.}
\label{fig:pyrochlore} 
\end{figure}

The pyrochlore lattice is a lattice of corner sharing tetrahedra, with four sublattices each of which occupies the sites of an FCC lattice. For a tetrahedron whose origin is at (0,0,0) the four sublattices (assigned labels 0,1,2,3) are located at ${\bf{r_0}} = \frac{1}{8}(+1,+1,+1) , {\bf{r_1}} = \frac{1}{8} (+1,-1,-1), {\bf{r_2}} = \frac{1}{8} (-1,+1,-1), {\bf{r_3}} = \frac{1}{8} (-1,-1,+1)$. Thus the line connecting sites 0 to 3 is the [110] direction, which is indicated in red in Fig.~\ref{fig:pyrochlore}. 
The 0 and 3 sites are collectively referred to as the $\alpha$ sites. 
The line connecting sites 1 and 2 is along the [1-1 0] direction (which is orthogonal to [110]), which is indicated in blue in Fig.~\ref{fig:pyrochlore}, these sites are collectively referred to as the $\beta$ sites.

We distinguish the global coordinate system, which we denote with $\{X,Y,Z\}$, from the local coordinate  
system at each rare earth site, which is given, following Ref.~\cite{huang2014}, as follows:
\begin{equation}
\begin{split}
\hat{z}_0&=\frac{1}{\sqrt{3}}(+1,+1,+1),~~\hat{y}_0=\frac{1}{\sqrt{2}}(0,+1,-1) ; \\
\hat{z}_1&=\frac{1}{\sqrt{3}}(+1,-1,-1),~~\hat{y}_1=\frac{1}{\sqrt{2}}(-1,0,-1) ; \\
\hat{z}_2&=\frac{1}{\sqrt{3}}(-1,+1,-1),~~\hat{y}_2=\frac{1}{\sqrt{2}}(-1,-1,0);  \\
\hat{z}_3&=\frac{1}{\sqrt{3}}(-1,-1,+1),~~\hat{y}_3=\frac{1}{\sqrt{2}}(-1,+1,0) .
\label{eq:zcoord}
\end{split}
\end{equation}  
Using a right-handed coordinate system, the local $\hat{x}$ axis is given by  $\hat{x}_i = \hat{y}_i \times \hat{z}_i$. It is this choice of local axes that gives the form of the Hamiltonian in Eq.~\eqref{eq:Hnn}. 

For a general magnetic field  $\mathbf{h}=(h_X,h_Y,h_Z)$ 
in the global frame, the local $x,\ y,\ z$ components are then the vector $\mathbf{h}$ projected along $\hat{x}_i, \hat{y}_i,$ and $\hat{z}_i$ axis, respectively, on site $i$:
\begin{equation}
    h^i_\alpha = \mathbf{h} \cdot \hat{\alpha}_i
\end{equation}	
Equation~\eqref{eq:Hnn} indicates that only the local $z$ components $h^i_z$ are of consequence in determining the coupling of the applied external magnetic field to a given pseudo-spin.

An interesting situation arises for a magnetic field of strength $h$ applied perfectly along [110], i.e., $h_X=h_Y=\frac{h}{\sqrt{2}}$ and $h_Z=0$ in the global basis. The Zeeman term is exactly zero for sublattices 1 and 2, i.e. the $\beta$ spins do not directly couple to the applied [110] magnetic field, for arbitrary field strengths. The Zeeman term is non-zero for sublattices 0 and 3, i.e., the $\alpha$ sites. For the case of large field strengths, the ground state corresponds to the pseudospin on sublattice 0 pointing ``out" and the one on sublattice 3 pointing ``in" i.e. along the local $z$ axes, away and towards the center of the tetrahedron, respectively. This configuration has been qualitatively depicted in Fig.~\ref{fig:pyrochlore}(b).

If the $\beta$ spins were truly free, the ground state would be exactly degenerate resulting in a residual entropy of $\frac{R}{2} \ln 2$ per site. In practice, the $\beta$ spins do couple to the $\alpha$ spins which in turn couple to the magnetic field, thus the $\beta$ spins are essentially free only at temperature scales above the scale of effective magnetic interactions. One should, however, expect to see the signatures of this effective decoupling in the specific heat and the associated magnetic entropy, especially at high fields where the dominant scale is set by the Zeeman term. Rather curiously, we find that our high-field experiments do not exactly adhere to this expectation, a finding we address further in the next sections.

For the case of [111] applied fields, by contrast, there is no such decoupling between sublattices, however sublattice 0 couples more strongly compared to sublattices 1,2,3. At high field strengths a unique ground state `3-in--1-out' (`1-in--3-out') configuration exists - this has been depicted in Fig.~\ref{fig:pyrochlore}(c).

\begin{figure}[htpb]
\includegraphics[width=0.53\textwidth]{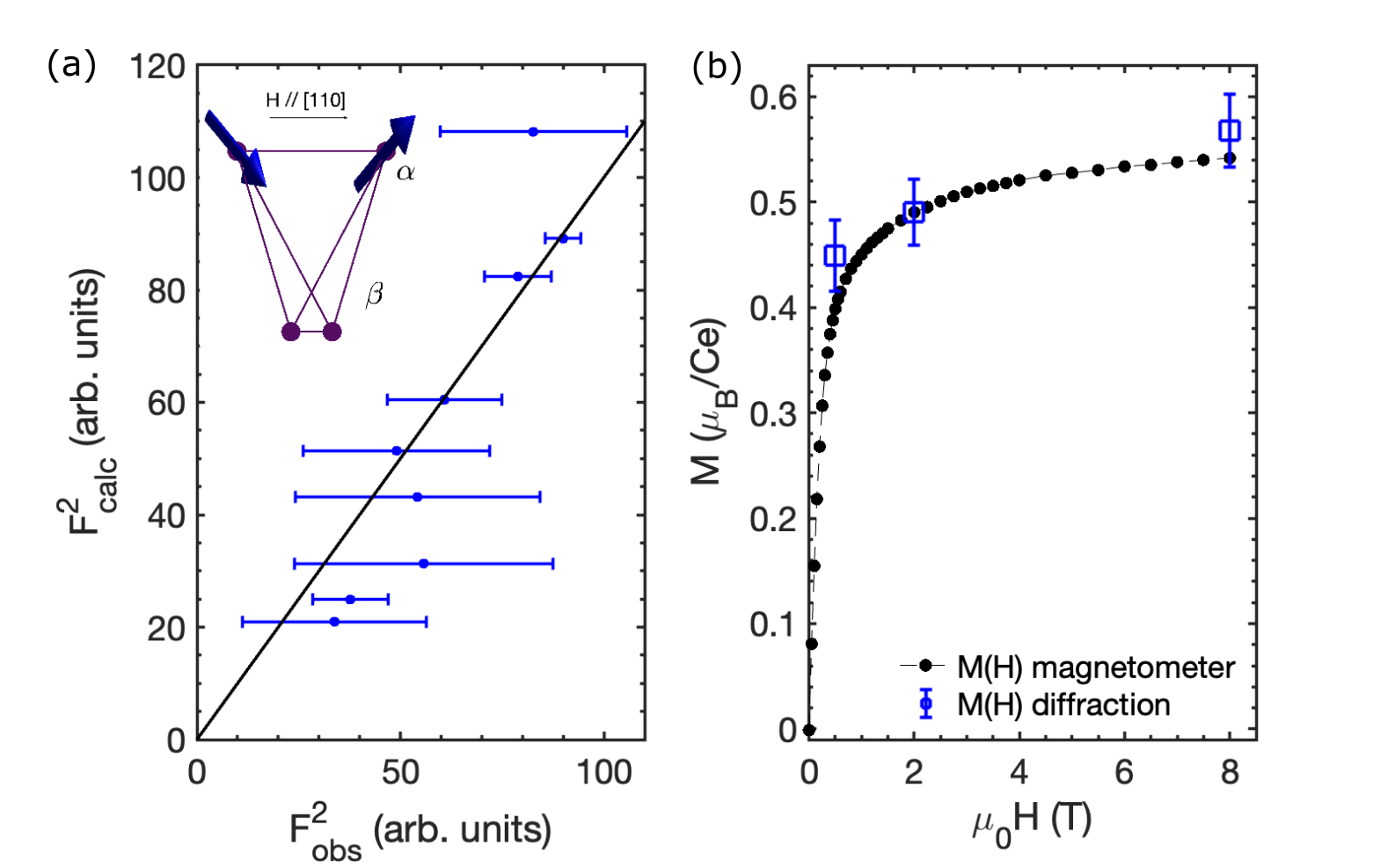}
\caption{(\textbf{a}) Observed \textit{vs} calculated intensities of magnetic Bragg reflections (8~T applied along the [110] direction at 0.05~K, agreement factor R$_f$ = 8.54). A scheme of the resulting magnetic structure on a single tetrahedron is depicted in the inset. (\textbf{b}) Comparison between the magnetization estimated from the neutron diffraction refinements at 0.05~K for various magnetic fields and the macroscopic magnetization curve measured at 0.08~K.}
\label{fig:neutron-fit}
\end{figure}

\section{Experimental results}
\label{sec:experiments}
We now present the details of the experiments performed, along with the key outcomes and puzzles they pose. In subsection~\ref{sec:neutron_110}, we show neutron diffraction data in [110] fields, which provide a picture consistent with the $\alpha$ spins ordering while the $\beta$ spins remain disordered. We then discuss in subsection~\ref{sec:spheat} specific heat results to be compared with numerical calculations presented in section ~\ref{sec:110_theory}. 

\subsection{Magnetic neutron diffraction and magnetization}
\label{sec:neutron_110}
The induced magnetic structure under application of a magnetic field parallel to the $[110]$ direction was investigated using single crystal neutron diffraction (Zebra, SINQ, PSI). The single crystal sample of \cehf~was cooled down using a dilution refrigerator inside a 10 T vertical cryomagnet. Reflections in and out of the horizontal scattering plane ($\pm15$ degrees) were collected using thermal neutrons (1.18~\AA) via a $^3$He point detector. The neutron beam was monochromatized and vertically focused using the (311) Bragg reflection of a germanium monochromator. Measurements took place at 1.5~K, 0.2~K and 0.05~K, under the application of a magnetic field $h$~=~0~T, 0.5~T, 2~T and 8~T. Magnetic Bragg intensities were extracted by subtraction of their respective nuclear contributions measured without magnetic field. All magnetic reflections could be accounted for using a $\textbf{k} = (0,0,0)$ propagation vector. Other directions of reciprocal space were investigated in order to check for magnetic Bragg scattering of a different propagation vector, which confirmed that the magnetic structure can be described within the chemical unit cell.

Fits of the experimental integrated intensities corrected for Lorentz factor were performed using the FullProf Suite program~\cite{FullProf_93} assuming the formation of $\alpha$ and $\beta$ chains and a 2-in-2-out local constraint (the fit is displayed in Fig.~\ref{fig:neutron-fit}), owing to the strong anisotropy and frustration experienced by the magnetic moments~\cite{PMB}. Across all temperatures and non-zero magnetic fields, the determined magnetic structures revealed a net magnetic moment on the $\alpha$ sites, lying along their respective local $[111]$ easy axis, polarized in the field direction and increasing with field strength. No net magnetic moment on the $\beta$ sites could be identified for $h$~=~0~T, 0.5~T, 2~T. Interestingly, for a magnetic field strength of $h$~=~8~T, allowing a small magnetic moment (0.10$\pm$0.05~$\mu_B$) parallel to the magnetic field leads to a slight improvement of the fit. This could be due to the increased number of fit parameters (and therefore be artificial) or could alternatively be real, potentially stemming from a slight misalignment of the field with respect to the ideal direction or due to a reduced single-ion anisotropy inherited from defective CEF environments~\cite{cehfo-poree}.

Magnetization \textit{vs} field was measured using SQUID magnetometers equipped with a miniature dilution refrigerator developed at the Institut N\'eel-CNRS Grenoble~\cite{Paulsen01}. We report data for the magnetic field applied along the [110] direction, as shown in Fig.~\ref{fig:neutron-fit}(b) for data at 0.08 K and in Fig.~\ref{fig:cv_lowfield}(d) at four representative temperatures. The magnetization of the sample was also calculated based on the magnetic structure determined from neutron diffraction at 0.05 K and compared with the macroscopic magnetization at 0.08 K (Fig.~\ref{fig:neutron-fit}(b)), showing a good agreement with, and further corroborating, the $\alpha$-chain/$\beta$-chain structure ansatz.

\subsection{Specific heat}
\label{sec:spheat}

A thin slice of \cehf~was prepared for specific heat measurement using the single crystal reported in \cite{cehfo-poree} and a tungsten wire saw. The [110] orientation of the slice's faces was assessed using a x-ray Laue diffractometer. Specific heat measurements between 0.05 K and 0.8 K were performed using a home-built calorimeter inserted in a dilution refrigerator at Université de Sherbrooke. For the rest of the paper, we will refer to this as ``experimental setup A". In this setup, the heater and the thermometer were glued directly to the sample using GE varnish. This assembly was suspended on zylon wires to decouple it thermally from the copper mount. Electrical contacts were made using 7 µm diameter superconducting NbTi wires. The thermal conductivity between the sample assembly and the copper mount was adjusted using a manganin wire, in order to have suitable relaxation times for the measurements. The manganin heat link was 1.7 mm long with a diameter of 0.1 mm. As well as providing the thermal link to the refrigerator, this short length of wire also provided significant structural support to the sample in addition to the very thin zylon filaments. The sample was oriented to apply the magnetic field along the [110] crystallographic direction. Data were acquired using the quasi-adiabatic heat pulse method.
 
Specific heat measurements between 0.4~K and 15~K were carried out using a Quantum Design physical properties measurement system (PPMS), with vertically applied magnetic fields up to 6~T. This is referred to as ``experimental setup B" for the rest of the paper. For this setup, the sample was mounted with its smoothest side facing down towards the puck's platform using an Apiezon grease, allowing the application of a magnetic field parallel to the [110] crystallographic direction. The platform itself is suspended by four wires, which also provide electrical connections to the heater and thermometer. Data measured on the nonmagnetic analog \lahf~were used to subtract lattice contributions. 

\begin{figure*}
\includegraphics[width=0.8\textwidth]{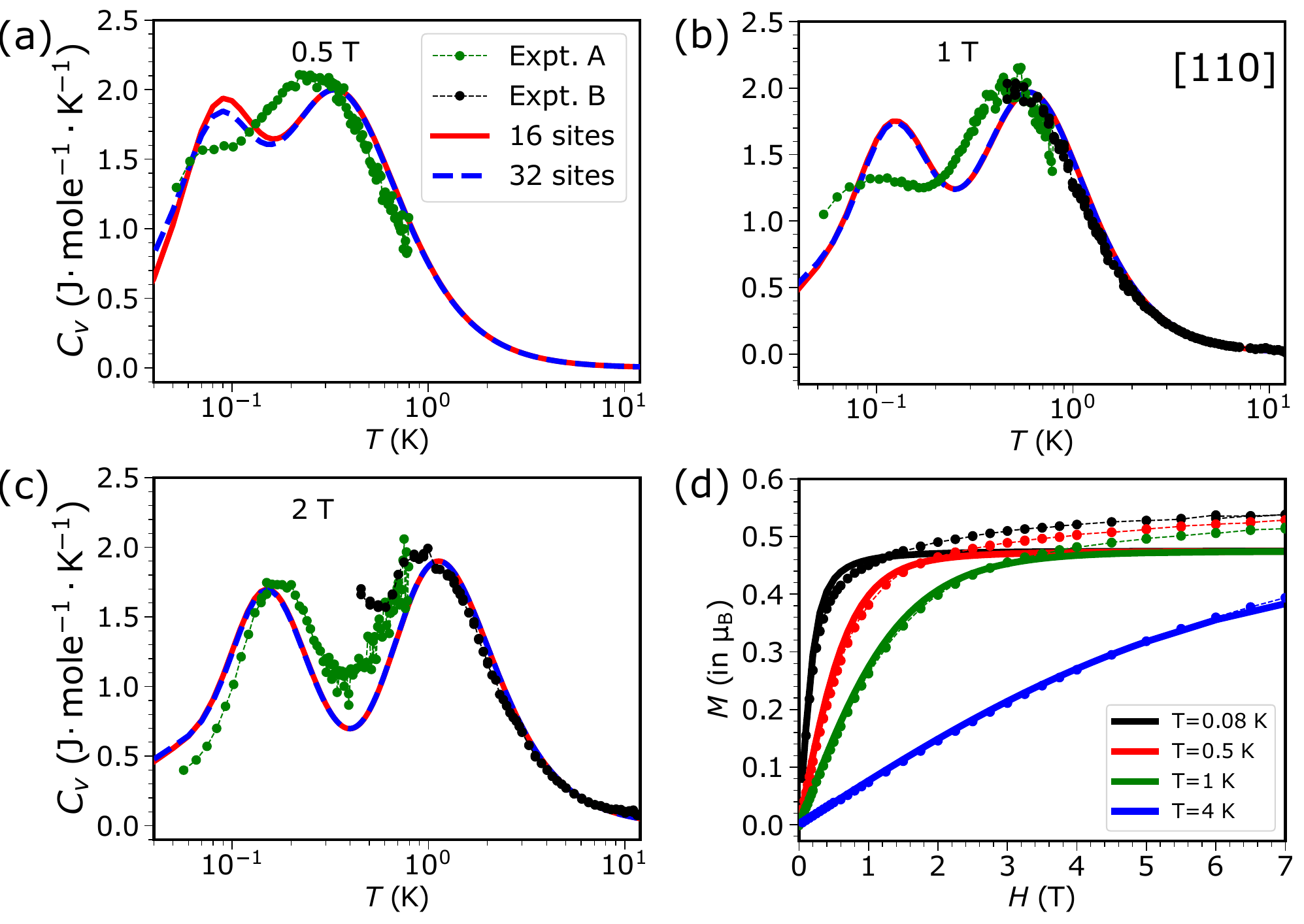}
\caption{(a-c) Comparison of experimental magnetic contribution to the specific heat data for ${\bf h}||$[110] (solid black/green circles) and calculated specific heat using FTLM performed on 16 sites (solid red) and 32 sites (dashed blue) for field strengths 0.5 T, 1 T and 2 T. The magnetic contribution (for setup B) was isolated by subtraction of the specific heat measured on the La$_{2}$Hf$_{2}$O$_{7}$ non-magnetic analog (for setup A, the lattice contribution was found negligible at all temperatures).  Figure (d) compares the bulk magnetization measured along [110] (solid circles connected by dashed line) with that obtained from FTLM (solid lines) using a 16-site cluster. FTLM calculations were performed with the model parameters in Eq.~\eqref{eqn:parameters}.
\ab{For the FTLM calculations on 16 sites $M=100$ Krylov vectors and $R=200$ seeds were employed. FTLM calculations on 32 sites were performed with $M=100$ and $R=5$ seeds for each of the eight momentum sectors~\cite{Changlani.YTO}.}
\label{fig:cv_lowfield}
}  
\end{figure*}

\begin{figure*}
\includegraphics[width=0.75\textwidth]{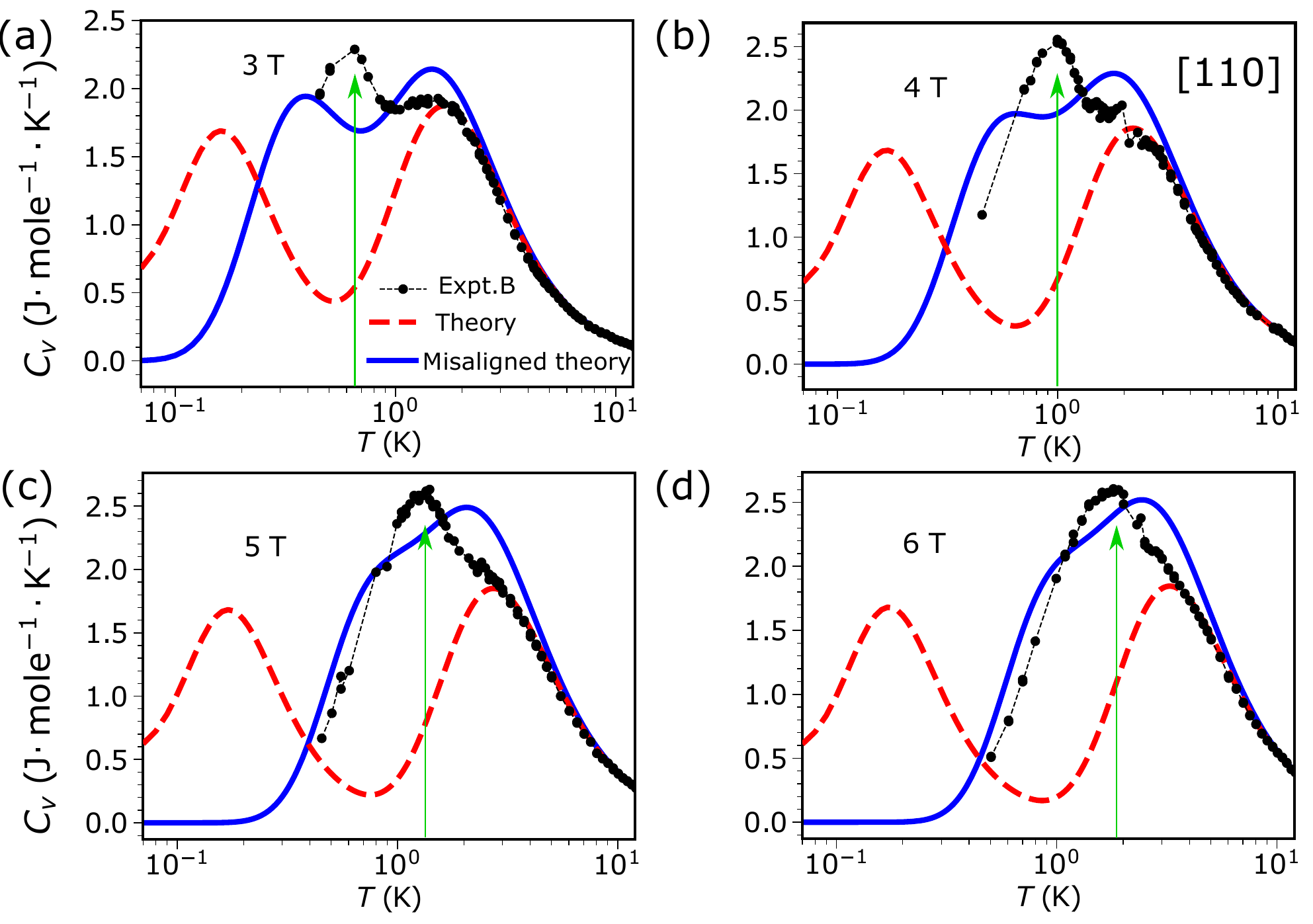}
\caption{
(a-d) Comparison of experimental magnetic contribution to the specific heat measured for $h\!\parallel\![110]$ (black solid circles), and calculated specific heat using FTLM 
with (solid blue line) and without (red dashed line) considering the misalignment. The magnetic contribution was isolated by subtraction of the specific heat measured on the La$_{2}$Hf$_{2}$O$_{7}$ non-magnetic analog. The vertical green arrow points at the low temperature peak of the experimental specific heat.  Calculations were performed using a 16-site cluster with the model parameters in Eq.~\eqref{eqn:parameters}.
}
\label{fig:cv_highfield}
\end{figure*}

Figures~\ref{fig:cv_lowfield}(a),(b),(c) and \ref{fig:cv_highfield} show experimental specific heat data for \cehf~ in a magnetic field along the [110] direction in both setup A (green) and B (black), for low ($h=0.5,1,2$ T) and high field strengths ($h=3-6$ T) respectively. Note that for the latter case, only data from setup B is available. The experimental results reveal a prominent two-step release of entropy, which appears as two outstanding peaks in the specific heat -- \ab{the lower temperature feature being somewhat sharp and the high temperature feature being a broader bump}. Note that such a two-step entropy release, if present, is not prominent for fields applied in the [111] direction \cite{PoreeArxiv2023}. The temperature of both peaks in the specific heat \emph{increases} with the applied field strength.

As emphasized earlier, the field applied along the [110] direction separates the spins in two types of chains, labeled $\alpha$ and $\beta$. Owing to the amplitude of the various effective interactions, the high-temperature bump arises from the Zeeman term of the $\alpha$ spins, scaling linearly with the field strength. The low temperature feature indicates correlations on the $\beta$-chains, as investigated in more detail in the next section.

\section{Theoretical analysis for [110] magnetic fields} 
\label{sec:110_theory}
We now compare our experimental results with the theoretical predictions based on our previously determined Hamiltonian parameters \cite{PoreeArxiv2023} reproduced in Eq.~\eqref{eqn:parameters}. These parameters were obtained from fitting quantum many-body calculations to thermodynamic measurements in [111] magnetic fields and rescaled classical Landau-Lifshitz spin/molecular dynamics~\cite{ncnf.zhang} to neutron scattering measurements in zero field. Since the data for [110] fields was not used for the fitting, a comparison between experiment and theory serves as a non-trivial check of the microscopic description. 

Theoretical results are obtained for both 16-site and 32-site pyrochlore clusters (with periodic boundary conditions) using the finite temperature Lanczos method (FTLM)~\cite{FTLM.Prelovsek} which has previously been applied for related systems~\cite{Changlani.YTO, Bhardwaj2022NatMat} (Further details of the method and its application to the pyrochlore geometry can be found in Ref.~\cite{FTLM.Prelovsek, Changlani.YTO}). While both cluster sizes are small, we see that their high temperature features coincide, a result expected when the correlation lengths are short (see red and blue curves in Fig.~\ref{fig:cv_lowfield} (a-c)). This good agreement also extends to lower temperatures, giving us confidence in the numerical results. \ab{Our calculations also suggest that finite size effects, which are minimally present at higher temperatures, do become more prominent at lower temperatures, especially at lower fields, see for example the results for 0.5 T in Fig.~\ref{fig:cv_lowfield}(a). This is anticipated given the existence of the putative gapless spin liquid at zero field~\cite{PoreeArxiv2023}. However, in practice even a modest field strength ($h \gtrsim$  1 T) appears to be dominated by the physics of a single tetrahedron, which is why we employ the 16-site cluster only for the rest of the paper.}

We also highlight the agreement of the computed specific heat with the high temperature portion of the experimental data, reaffirming faith in the theoretical microscopic description and the parameter values, specifically 
suggesting that the $g$ value is well-constrained. 
At lower temperature, however, there are quantitative differences with respect to the experiment. Calculations indeed predict two peaks in the specific heat (see Fig.~\ref{fig:cv_lowfield} (a-c) and dashed red curves in Figure \ref{fig:cv_highfield}), but the field evolution of the low temperature peak is not correctly captured. 

We also compare experimental magnetization curves with calculations based on parameters in Eq. \ref{eqn:parameters} (see Fig.~\ref{fig:cv_lowfield}(d)). Our theoretical curves (computed for the 16 site pyrochlore cluster) agree well with the experiment for low fields at all temperatures, and for all fields at higher temperatures. At high fields and especially low temperatures, prominent deviations are observed, notably, the lack of saturation of the experimental magnetization. As mentioned above, this regime is expected to be governed by the physics of a single tetrahedron, thereby ruling out significant finite size effects, with interactions playing only a diminished and subdominant role. Thus it becomes imperative to scrutinize the origin of this discrepancy.
 
We revisit the temperature dependence of the specific heat in a [110] magnetic field. For large fields, the location of the (high temperature) Schottky bump 
originates from the $\alpha$ spins that align themselves to minimize their Zeeman energy. As the magnetic field strength is increased, the magnetic energy contribution from the $\alpha$ chains increases and hence a higher temperature is required to completely disorder these spins, which is associated with the destruction of short-range order. However, this does not address the underlying reason for the existence of the low temperature feature and its movement with increasing magnetic field. To address this we investigated three scenarios, including one based on existing literature, 
which we discuss next. 

\subsection{First scenario: Indirect coupling of \texorpdfstring{$\beta$}~ spins to a [110] magnetic field mediated by the \texorpdfstring{$\alpha$}~ spins} 
\label{sec:first_scenario}
The $\beta$ spins couple to one another by field-dependent effective interactions, this is mediated via their interaction with the $\alpha$ chains. This was placed in a theoretical framework by Placke, Benton and Moessner (PMB)~\cite{PMB} using a real space perturbation theory approach and order by disorder arguments~\cite{Henley_obd} treating each $\beta$ chain as an effective Ising degree of freedom. For dominant $J_y$ and small $J_x,J_z$ while keeping the [110] field large, PMB calculated the couplings between nearest neighbor Ising spins to be,
\begin{eqnarray}
\label{eq:PMB-pert}
J_{ch} &\propto& J_x^2 J_z^2 \frac{(J_x-J_z)^2}{J_y^5}\Gamma_y(h) \ 
\end{eqnarray}
where $\Gamma_y(h)$ decays as $h^{-2}$ as $ h \rightarrow \infty$. 
This functional dependence suggests that the order-disorder transition of the $\beta$ chains should occur at \emph{lower} temperatures upon increasing the field strength. This is \emph{opposite} to the trend we observe in experiments, and thus we see no evidence of the PMB scenario in our current experiments. We discuss this further in Appendix~\ref{app:first_scenario}.

\subsection{Second scenario: Higher-order Zeeman coupling of $\beta$ spins to [110] field}
\label{sec:second_scenario}
We also explored the possibility of 
a direct coupling of the $\beta$ spins with the applied [110] magnetic field by incorporating the next order Zeeman term~\cite{Patri-octupolarQSI}. 
This is allowed by symmetry for the projected effective Hamiltonian, restricted to the doublet subspace, 
and the term is given by
\begin{equation}
\label{eq:Hz}
\begin{split}
H^i_\text{eff-Z}&=-\mu_Bh_z^i(g_xs_x+g_zs_z) 
-g_y\mu_B^3h_y^i((h_y^{i})^2-3(h_x^i)^2)s_y \\
&-\overset{\sim}{g_x}\mu_B^3h_x^i((h_x^{i})^2-3(h_y^i)^2)s_x. 
\end{split}
\end{equation}
When the field is along [110] direction, $h_x^1=-\frac{\sqrt{3}h}{2}$, $h_y^1=-\frac{h}{2}$ and $h_x^2=0$ (where $h$ is the field strength) and this makes the coefficient of $\overset{\sim}{g_x}$ in the above equation zero for sublattices 1 and 2. 
In Appendix~\ref{app:second_scenario} we discuss the details of our estimation of $g_y$ from ab-initio (CEF) calculations. We find it to be small $ \approx 0.0004$ meV$^{-2}$, this smallness rules out this mechanism as a quantitative explanation of the two step release of entropy in \cehf.

 \begin{figure*}
\includegraphics[width=0.98\textwidth]{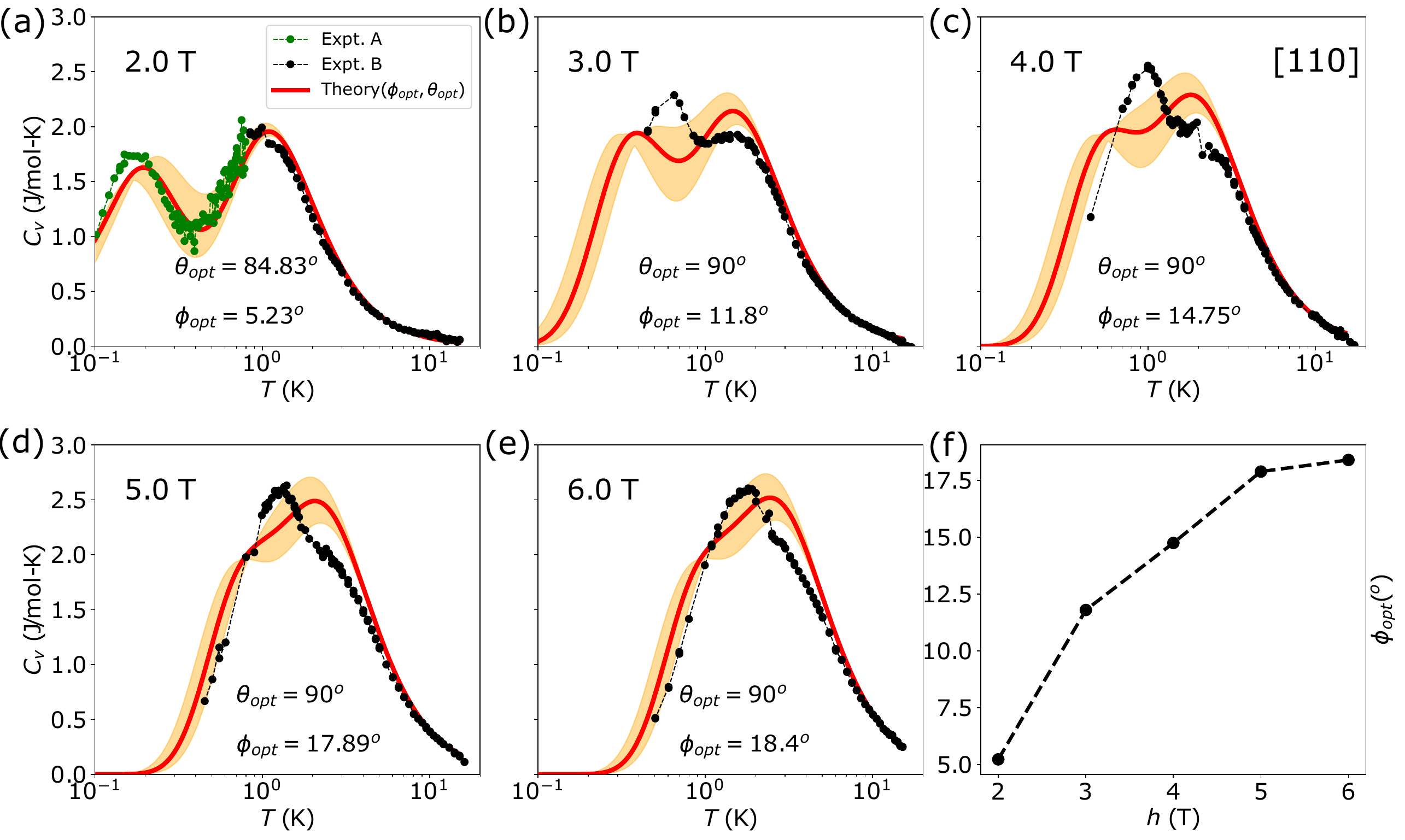}
\caption{Panels (a)-(e) show the calculated specific heat, for representative field strengths, when the field is not perfectly aligned along the [110] direction. The red solid line in each panel represents the specific heat obtained by using the derived optimum values of the deviation angles $\phi_{opt}$ and $\theta_{opt}$ for each value of field strength. The orange region in each panel shows the variation of the specific heat as the deviation angle $\phi$ is varied from $\phi_{opt}-3$ degrees to $\phi_{opt}+3$ degrees in increments of one degree while the deviation angle $\theta$ is fixed at the optimum value for a given value of the field strength. The experimental magnetic contribution to the specific heat data with the field along [110] is shown in black points. Panel (f) shows the variation of the deviation angle $\phi_{opt}$ with the field strength. All calculations were performed using 16 sites with the model parameters in Eq.~\eqref{eqn:parameters}.
}       
\label{fig:cv_maln}
\end{figure*}

\subsection{Third Scenario: Misalignment of the sample away from [110] and physical reason for two step release of entropy}
\label{sec:third_scenario}
The third scenario involves a direct Zeeman coupling  of the $\beta$ chains to the magnetic field by assuming its misalignment away from the [110] direction. Since the  preparation of specific heat measurements involves several steps subject to uncertainties (alignment from Laue diffraction images, sample cut, sample mounting on a given specific heat setup, and alignment of that setup relative to the superconducting solenoid), a slight misalignment of the applied magnetic field with respect to the required crystallographic orientation can hardly be avoided already at the start of measurements.
Added to that, the torque induced by the application of a magnetic field on the sample may induce a deviation from the starting geometry. In particular, according to our experience, the measurement platform of the specific heat puck (commonly used for measurements in the PPMS) can tilt by several tens of degrees at low temperature, without breaking the wires, when applying the field along the hard axis of a magnetic sample.

We now quantitatively determine the effect of misalignment on the specific heat data. To do so, we parameterize the global-basis components of a magnetic field tilted away from the [110] direction, as a function of its strength $h$, polar angle $\theta$, and azimuthal angle $\phi$ (see Fig.~\ref{fig:pyrochlore}),
\begin{subequations}
\begin{eqnarray}
\label{eq:mis1}
h_X&=&\frac{h}{\sqrt{2}}(\sin \theta \cos \phi + \sin \theta \sin \phi)\\
\label{eq:mis2}
h_Y&=&\frac{h}{\sqrt{2}}(\sin \theta \cos \phi - \sin \theta \sin \phi)\\
\label{eq:mis3}
h_Z&=&h \cos \theta 
\end{eqnarray}
\label{eq:theta-phi}
\end{subequations}
\noindent{where we have taken $\theta$ to be the angle with respect to the $Z$ axis, and $\phi$ to be the angle with respect to the $45$ degree line in the $X-Y$ plane. The case of $\theta=90^\circ$ and $\phi=0^\circ$ corresponds to the perfect alignment along the [110] direction. Due to the cubic symmetry of the pyrochlore lattice, our results for the specific heat for a given $\theta,\phi$ will be identical to those for $\theta,-\phi$.}

Our attempts to explain the data with a single set of misalignment angles were not successful, so for each magnetic field strength $h$ we optimize the values of $\theta$ and $\phi$ independently to best fit the experimental [110] specific heat data. For each $h$, the initial guess of $\theta=88^\circ$ and $\phi=10^\circ$ was chosen with $\theta$ and $\phi$ constrained between 45$^{\circ}$ to 90$^{\circ}$ and 0$^{\circ}$ to 40$^{\circ}$ respectively. At each stage of the optimization (using the SLSQP algorithm in python) an FTLM calculation with $R=100$ seeds and $M=50$ (where $R$ is the number of independent Lanczos runs and $M$ is the dimension of the Krylov space) was performed\ab{, which were found to be sufficient for attaining convergence within the temperature range considered}. The cost function was designed to minimize the difference between the experimental and FTLM data. 

The results of our specific heat curves for the optimized misalignment angles $\theta_\text{opt}(h)$ and $\phi_\text{opt}(h)$ are shown in Fig.~\ref{fig:cv_maln}(a-e). Our optimized values of $\theta_\text{opt}$ and $\phi_\text{opt}$ suggest that the specific heat is more sensitive to values of $\phi$ rather than $\theta$: the latter is found to be only a few degrees off from 90 degrees (the fits that allowed $\theta$ to exceed 90 degrees also found the optimized value to be close to 90 degrees). We have thus shown the extent of variation of specific heat in Fig.~\ref{fig:cv_maln} for the cases where $\theta_\text{opt}(h)$ was fixed and $\phi$ in the range $ \phi_\text{opt}(h) - 3^\circ \le \phi \le \phi_\text{opt}(h) + 3^\circ$. The optimal curves are shown in red and all the curves for different values of $\phi$ accurately capture the high temperature tail. This is because at these temperatures the spins are disordered and tilting the field further does little to change the situation. Importantly, however, the different curves show variation in their lower temperature features based on which we infer that our $\phi_\text{opt}$ has an error bar of a few degrees. 

Owing to the sub-meV interaction scales, fields strengths of even a few Tesla 
are large 
enough for spin-spin interactions to be a sub-leading effect. For example, at $6 $T there are almost no finite size errors as the thermodynamics is largely governed by the response of four (almost) non-interacting sublattices. However, our optimized FTLM and experimental data do not perfectly match in this regime suggesting that effects beyond those incorporated in our model are required to better reproduce the experimental data.

\begin{figure}
\includegraphics[width=0.475\textwidth]{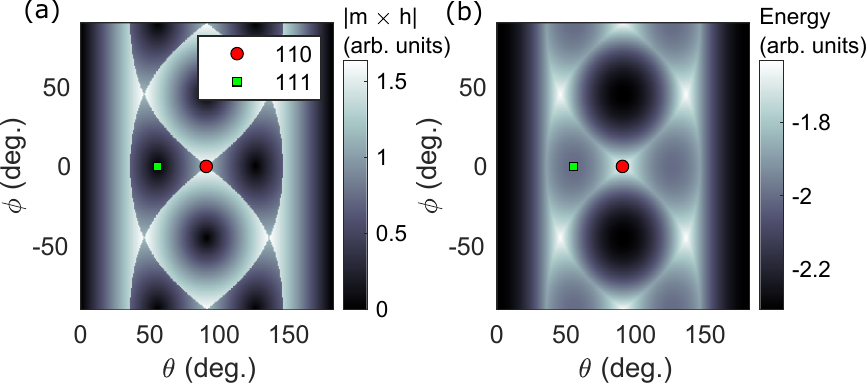}
\caption{Expected torque for classical Ising moments on a tetrahedron, without interactions. The moment configuration minimizing the Zeeman energy for each field direction was used to evaluate the torque. Panels (a) and (b) illustrates the magnitude of the torque and the Zeeman energy, respectively, as function of the field direction. For field along [111], the system sits at a local minimum with zero torque. Any misalignment from this direction leads to a torque bringing the system back to the [111] minimum. For field along [110], the system experiences a large torque and is unstable, as it sits at a local maximum.}
\label{fig:torque}
\end{figure}

The qualitative trend from our fits suggests $\theta_\text{opt}$ remains close to $90^{\circ}$ while $\phi_\text{opt}$ increases with field strength. The increase of $\phi_\text{opt}$ is steeper at low field and tends to saturate above 5 T, as seen in Fig.~\ref{fig:cv_maln}(f). 
To understand this behavior, we have calculated the torque expected for classical Ising moments on a tetrahedron as a function of field direction. The results are presented in Fig.~\ref{fig:torque} and show that the torque has a maximum for the [110] direction, around which it varies very differently depending on the direction of approach, $\theta$ or $\phi$. Going away from the [110] direction in either $\theta$ or $\phi$ goes towards a minimum in torque corresponding to a lower Zeeman energy (Fig.~\ref{fig:torque}\text{(b)}). This energy landscape rationalizes the increase in $\phi_\text{opt}$ deduced from the fits of the specific heat data.
From an experimental viewpoint, both measurement setups for specific heat rely on thin wires to hold the sample, or sample platform, in the starting orientation, thus leaving room for movements under applied magnetic field. Although the fitted $\phi_\text{opt}$ values from the high-field PPMS data appear relatively large, they are broadly consistent with our observations of widely tilted platform with other samples (see Appendix~\ref{App:B}).

The emergent picture of the two peak feature in the specific heat in a misaligned sample is as follows. At low temperature and high $[110]$ fields, the $\alpha$ spins choose an axis that is (almost) the local +$z$ or -$z$ direction. Because of this locking and because the $\beta$ spins now see a small but direct field and also interact with the $\alpha$ spins, they too choose a unique ground state from an extensively degenerate manifold. Once the temperature is increased, the $\beta$ spins disorder first, yielding the first peak. Upon increasing the temperature further, the $\alpha$ spins disorder, which corresponds to the second bump. These scales are distinct at low field strengths where the misalignment is small. At high field strengths the misalignment is larger, and the Zeeman energy scales for $\alpha$ and $\beta$ spins become comparable. The double peak feature thus gradually fades away to yield a specific heat with a single bump as the field is increased, an effect that is indeed captured by our FTLM simulations.

\begin{figure}
\includegraphics[width=0.95\linewidth]{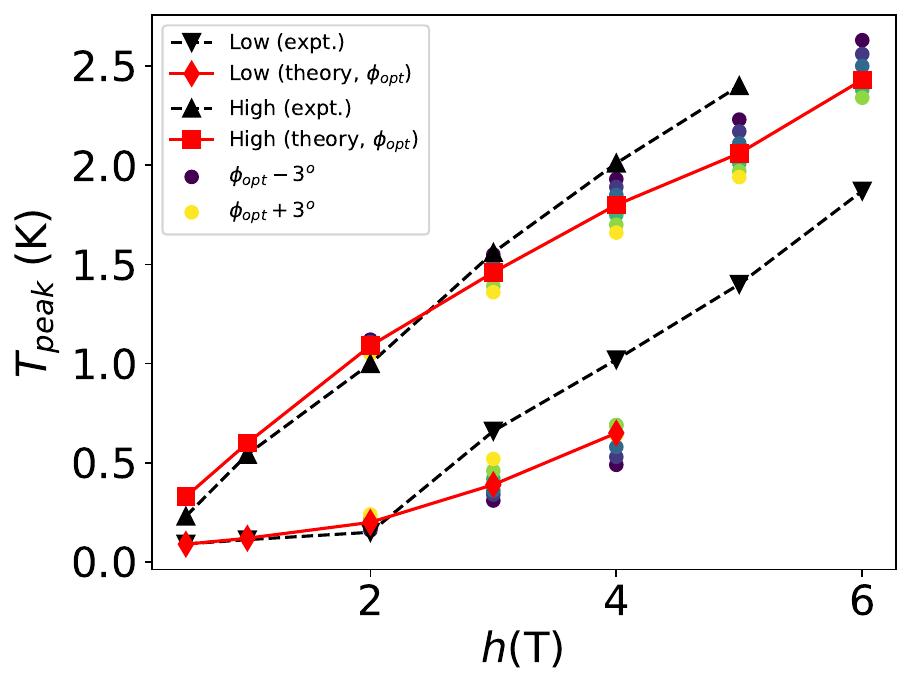}
\caption{Field strength dependence of the low and high temperature peaks in specific heat for field along the [110] direction. The low temperature peaks at 0.5 T, 1 T and 2 T were extracted from data measured using the experimental setup A. The misalignment at 0.5 T and 1 T is not significant and hence the calculated positions of both low and high temperature peaks at these two field strengths were obtained by assuming perfect alignment of the field along the [110] direction. FTLM was used to evaluate all the specific heat curves using 16 sites with the model parameters in Eq.~\eqref{eqn:parameters}. \ab{Data for several misalignment angles in the range $\phi_{opt} \pm 3^{\circ}$ are also shown.
}}
\label{fig:peakloc}
\end{figure}

In Fig.~\ref{fig:peakloc} we quantitatively check our theoretical results for the two step release of entropy with respect to the experimental results. The experimental results (where the peak could be clearly discerned) suggest that both peak locations move linearly with field strength. This broadly supports the assertion that the $\beta$ sublattice also experiences a direct magnetic field in accordance with the misalignment theory. The location of the high-temperature bump calculated with FTLM is in excellent agreement with experiments, but the low temperature feature, while showing a generally correct qualitative trend, displays some quantitative discrepancy. \ab{We also note in Fig.~\ref{fig:peakloc} that the slope of the low-temperature peak changes abruptly past 2 T, which is due to data from two different setups (setup A at low fields and setup B at higher fields), 
but no such difference 
is present in the high-temperature bump. We attribute this to the location of the high temperature bump being less susceptible to misalignment (which we expect to be different in the two setups), as it arises from melting of the $\alpha$ spins which directly couple to the [110] magnetic field.}

Lastly, we also explore the effect of misalignment on the magnetization. 
Note that the experimental magnetization has a gradual, but non zero, slope at large values of field strength, which is a regime primarily dominated by physics of a single tetrahedron. 
Further investigations were carried out to probe the role of impurities in the sample - we found that their effects are likely too small to explain the magnitude of the feature seen in the experiment. (We refer the reader to Appendix~\ref{App:C} for 
discussion on the role of impurities.)

Calculations from \textit{ab-initio} calculations (using PyCEF~\cite{Scheie_PYCEF} and SPECTRE~\cite{SPECTRE}) working in a Hilbert space of 14 states, i.e. keeping all the $J=5/2$ and $J=7/2$ states, yielded a small but non-negligible van Vleck contribution to the susceptibility, $\chi_0 \approx 2.1 \times 10^{-3} \mu_B/T$. This non-zero contribution emerges from the presence of higher crystal field levels beyond the doublet, and results in a non-zero slope in the $M$ versus $h$ curve at large fields. (This latter feature does not exist in the pseudo spin-1/2 model with only a linear Zeeman term.) Despite its qualitative correctness, quantitatively the calculated $\chi_0$ is a factor of 5 smaller than what is needed to explain the slope seen in the magnetization experiment. 

\begin{figure}
\includegraphics[width=0.95\linewidth]{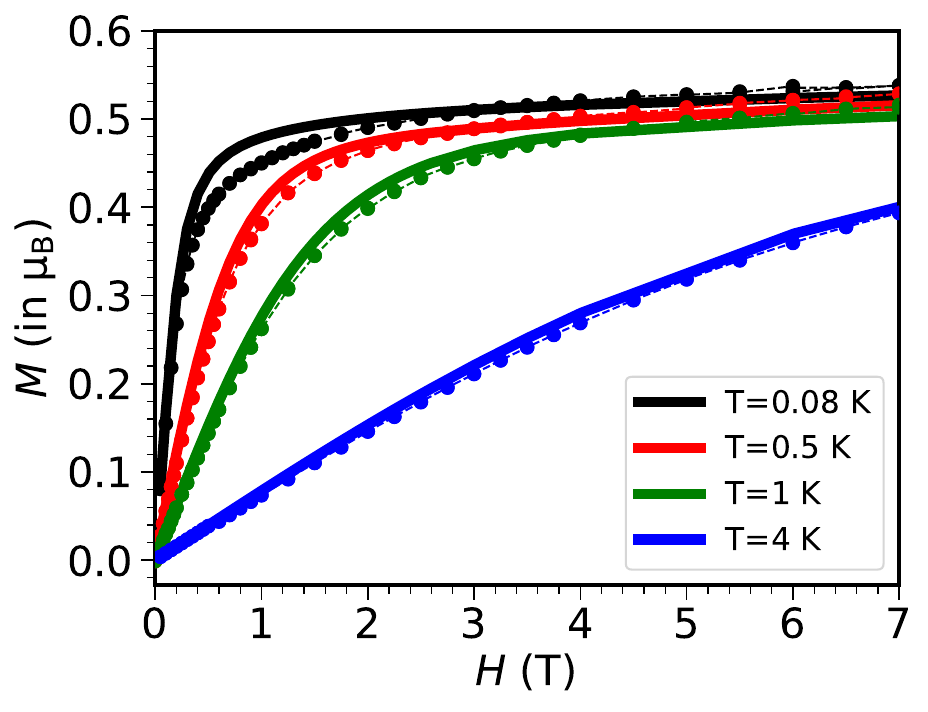}
\caption{
Calculated magnetization (solid lines) assuming a constant misalignment angle of 5 degrees and the added correction from the van Vleck contribution. The solid dots connected by dashed line represent the [110] experimental magnetization. FTLM was used to evaluate all the magnetization curves using 16 sites with the model parameters in Eq.~\eqref{eqn:parameters}.
}
\label{fig:mag-misaligned}
\end{figure}

To address these inadequacies, we model the magnetization setup using only a constant, and small, misalignment angle $\phi$ of $5$ degrees keeping $\theta$ fixed at 90 degrees. (Unlike the specific heat setup where the misalignment angle changes with the field strength, the experimental magnetization measurement setup is not expected to exhibit any such field dependence, motivating our choice of the misalignment model.) Fig.~\ref{fig:mag-misaligned} shows our FTLM results for the magnetization by incorporating both the van Vleck and misalignment contributions. The results, especially at high fields, are greatly improved with respect to those presented in Fig.~\ref{fig:cv_lowfield}(d), at the cost of only a small disagreement at low fields and low temperatures. 


\begin{figure*}
\includegraphics[width=0.95\textwidth]{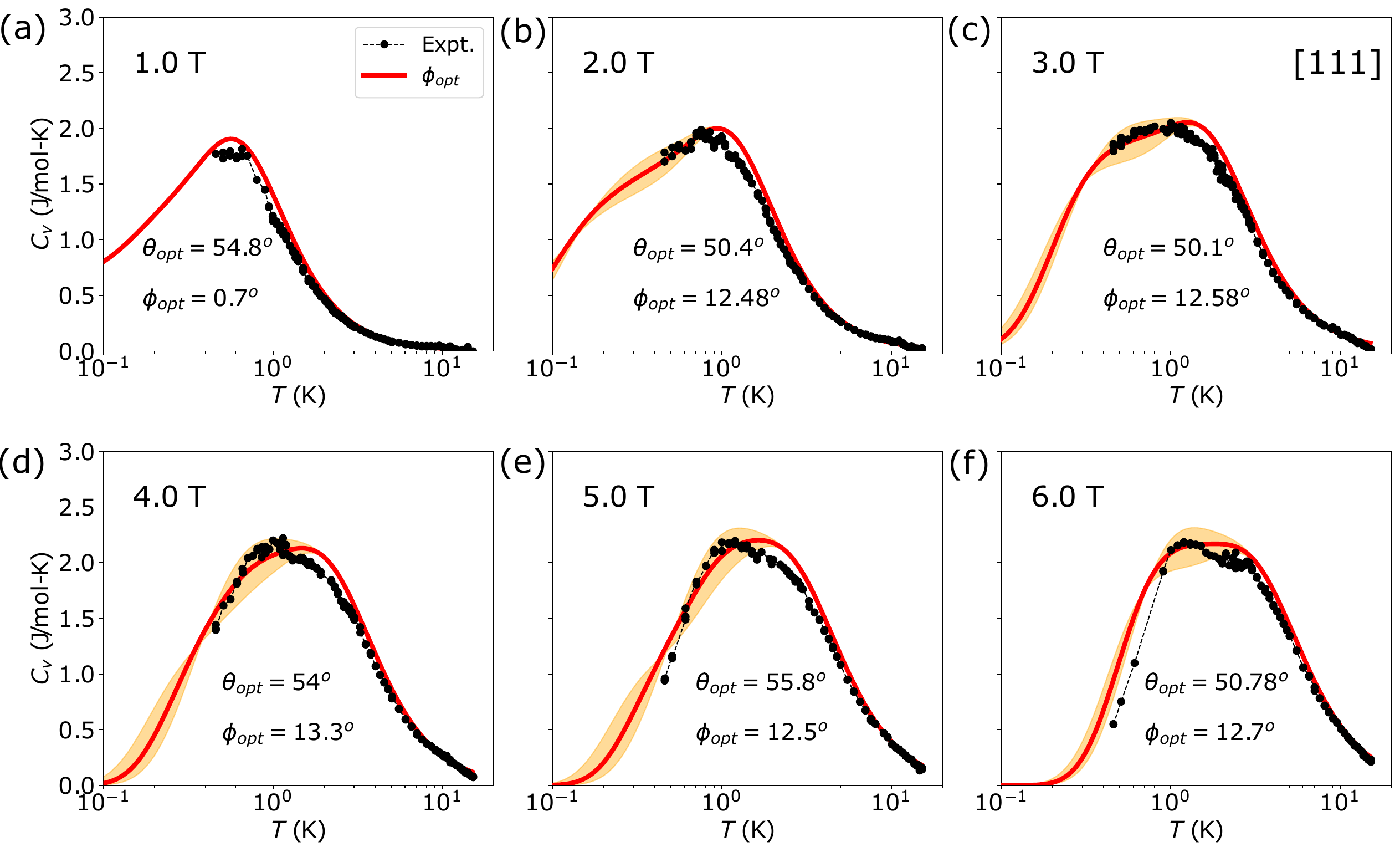}
\caption{Calculated specific heat when the magnetic field is not perfectly aligned along the [111] direction. The red solid line in each panel represents the specific heat obtained by using the derived optimum values of the deviation angles $\phi_{opt}$ and $\theta_{opt}$ for each value of field strength. The orange region in each panel shows the variation of the specific heat as the deviation angle $\phi$ is varied from $\phi_{opt}-3$ degrees to $\phi_{opt}+3$ degrees in increments of one degree while the deviation angle $\theta$ is fixed at the optimum value for a given field strength. The experimental magnetic contribution to the specific heat data with the field along [111] is shown as black points. The experimental data are taken from Ref.~\cite{PoreeArxiv2023} and FTLM calculations were performed using 16 sites with the model parameters in Eq.~\eqref{eqn:parameters}. 
\label{fig:cv_maln_111}
}    
\end{figure*}


\section{Thermodynamics and effects of misalignment for magnetic fields applied along the [111] direction}
\label{sec:111}
For completeness, we have revisited the specific heat data for fields applied along the [111] direction \cite{PoreeArxiv2023}. This data was instrumental in obtaining the optimal Hamiltonian parameters in prior work, and was very well reproduced by the theory even without introducing a misalignment. For this field direction, in the absence of any misalignment, all four sublattices couple to the field although with different effective strengths - sublattice 0 couples most strongly while sublattices 1,2,3 have equivalent strengths. 

At low fields and low temperatures we expect the presence of a putative U(1) QSL -- the evidence being based on the lack of magnetic ordering, e.g., the absence of any Bragg peaks in neutron scattering, and the observation of a continuum of spin excitations instead~\cite{PoreeArxiv2023}. On heating, such a state evolves into the disordered paramagnetic state by a crossover, potentially explaining why there is only a single bump in the specific heat on a scale that depends both on the $J$'s and $h$. 

At higher fields, the [111] magnetic field selects a unique symmetry broken state  `3-in--1-out' (`1-in--3-out') state.  It too is related to the disordered state at high temperature only via a crossover. However, on looking closely at the data, we observe two fused features in the specific heat at high fields - this can be attributed to the coupling of the field to sublattice 0 being inequivalent to that of sublattices 1,2,3. 

We now investigate the effect of misalignment of [111] fields. We parameterize the field components for the misalignment using Eq.~\eqref{eq:theta-phi} where $\theta$ and $\phi$ represent the deviation angles defined previously in Fig.~\ref{fig:pyrochlore}. Here, the case of $\theta\approx 54.7^\circ$  and $\phi=0^\circ$ represents perfect alignment along the [111] direction.

The role of misalignment, while still important in controlling the height of the specific heat peaks, in general, is greatly diminished compared to the case of the [110] field. 
We have fitted the misalignment angles $\theta$ and $\phi$ for the case of specific heat in a [111] field using a fitting procedure similar to the case of a [110] field. In Fig.~\ref{fig:cv_maln_111} we plot the specific heat profiles, computed with FTLM, using these optimum values of misalignment angles. Even though the fitting procedure results in a misalignment of about 10 degrees, there is very little to distinguish the results with those for the case of no misalignment (0 degrees) i.e. the model does not improve significantly the global agreement with the data. This observation is consistent with torque calculations that indicate that a field applied close to the [111] direction is expected to realign this sample direction along the field, as illustrated in Fig.~\ref{fig:torque}.

\section{Conclusions}
\label{sec:conclusion}
Dipole-octupole systems serve as a rich source of pseudospin-1/2 physics and offer access to phases where quantum fluctuations are enhanced by geometric frustration, an amalgamation of conditions that are conducive for the occurrence of quantum spin liquidity \cite{huang2014,GangChen2017PhysRevB,PMB,Patri-octupolarQSI}. In contrast to the familiar dipolar systems, the relative inertness of the octupolar components to applied magnetic fields has profound consequences, and this paper explored a set of unusual effects emerging from such physics. 
While the current paper focused primarily on previously unpublished experimental data for \cehf, we believe that most conclusions should also transfer to its sister compound \cezr\ \cite{CZO.Gao, CZO.Gaulin,Bhardwaj2022NatMat}. 

Our experiments on \cehf~(neutron diffraction, specific heat measured with two independent setups and magnetization) and corresponding numerical simulations agree remarkably well with each other in the low field limit, providing confidence in the microscopic description and our previously determined parameters~\cite{PoreeArxiv2023}. However, data in the high field limit, which in principle should be the easiest to explain, pose puzzles which we have addressed here. In particular, our experimental data for the specific heat of \cehf~exhibit two distinct peaks indicating a two-step release of entropy. 
\ab{ The low temperature peak could be associated with a phase transition, as has been suggested recently~\cite{sanders2024do, ybk_felix_phase} and whose details remain to be investigated further-- our current numerical analysis suggests that its observed movement with field is primarily due to Zeeman coupling of the field to $\beta$ chains which is otherwise forbidden in idealistic settings.} A more refined look at the experiments and theory was required to decipher the origins of our experimental observations. 

We have explored three scenarios which would make the $\beta$ chains directly visible to the applied [110] field. The first explores the effective interaction between the $\beta$ chains mediated by the $\alpha$ spins locked to the [110] field, which emerges from an order-by-disorder mechanism proposed by Ref.~\cite{PMB}. 
We deduce that the theory would predict a low temperature specific heat feature that \textit{decreases} with field strength, rather than the \textit{increase} seen in our experiments. Second, we investigated the possible importance of symmetry-allowed terms beyond the linear Zeeman term, in principle, a cubic Zeeman term directly couples $\beta$ spins to the field. 
Using CEF and spin-orbit coupled \textit{ab-initio} calculations we find that this effect is too small. 

This leaves us with a third scenario that enables the $\beta$ chains to see the [110] field directly: a slight misalignment of the magnetic field from the [110] direction. 
But how large is this effect? The answer to this question led us to a systematic comparison between the misalignment theory, whose effects were calculated with FTLM, and the experiments. We estimated the misalignment angles which were found to increase with the field strength, consistent with a simple physical model of the experiment. The specific heat curves obtained with the misaligned field match reasonably well with the experimental data and thus we attribute the observed scaling of the low temperature peak with the applied field strength to the misalignment. This understanding led us to review our data for [111] fields, where we found that misalignment does not significantly improve the global agreement with the data in this case, as expected from the calculated directional dependence of the magnetic torque for dipoles on a pyrochlore lattice.

We conclude by noting that our work benefited from a close synergy between experiment and theory, the latter employed both \textit{ab-initio} and many-body calculations. Given that multiple competing scales exist in multipolar magnetic systems, careful comparisons are indispensable for determining the effective microscopic description~\cite{PoreeArxiv2023} and also identifying the origins of the observed responses in applied magnetic fields. More generally, many of the strategies employed in this work should be transferable to other magnetic materials (in the dipole-octupole family and beyond), enabling a better qualitative and quantitative understanding of their physical properties.

\section{Acknowledgments}
We thank O. Tchernyshyov, R. Moessner, B. Gaulin, Y. Kim, S. Zhang and A. Patri for insightful discussions and correspondence. We also thank A. Scheie for help with the PyCEF software~\cite{Scheie_PYCEF}. H.J.C. was supported by National Science Foundation CAREER grant No. DMR-2046570. A.B. and H.J.C. thank Florida State University and the National High Magnetic Field Laboratory for support. The National High Magnetic Field Laboratory is supported by the National Science Foundation through NSF/DMR-2128556 and the state of Florida. We also thank the Research Computing Center (RCC) and Planck cluster at Florida State University for computing resources. H.Y. and A.H.N. were supported by the National Science Foundation Division of Materials Research under the Award DMR-1917511. The theoretical analysis by A.H.N. was supported by the U.S. Department of Energy, Office of Basic Energy Sciences under Award no. DE-SC0025047. N.G. and J.A.Q. acknowledge the support of the Canada First Research Excellence Fund (CFREF). We acknowledge funding from the Swiss National Science Foundation (project No. 200021\_179150) and European Commission under grant agreement no. 824109 European `Microkelvin Platform'. This work is also based on experiments performed at the Swiss spallation neutron source SINQ (Paul Scherrer Institute, Switzerland)

\begin{figure}
\includegraphics[width=0.43\textwidth]{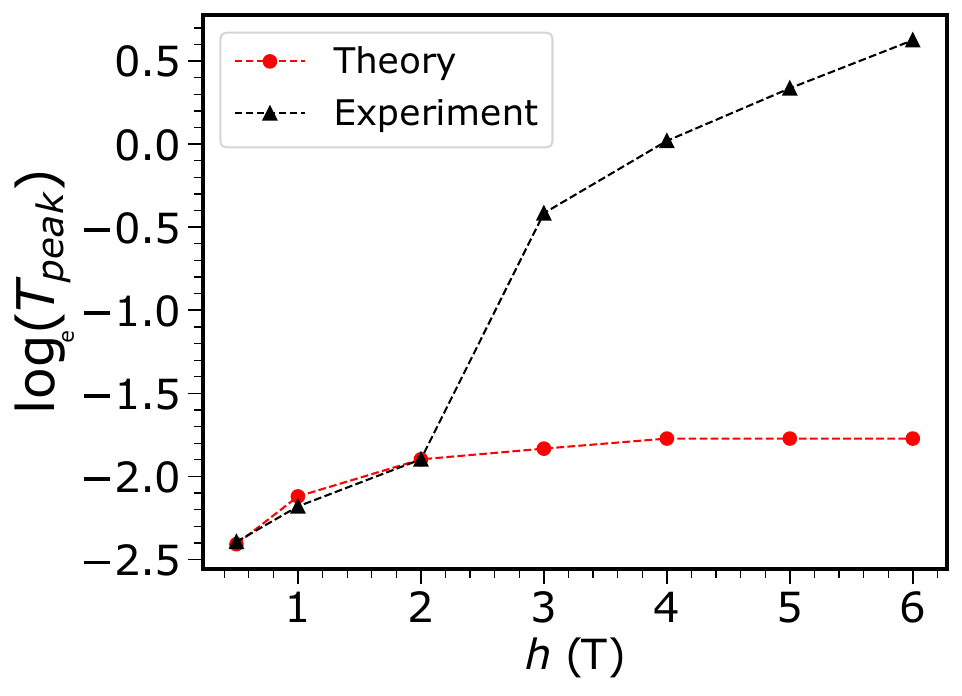}
\caption{
Field strength dependence of the log$_e$ of the low temperature peak of the specific heat curve obtained from FTLM calculations (assuming perfect alignment) and experimental data for [110] magnetic fields. Calculations were performed on a 16-site cluster with the model parameters in Eq.~\eqref{eqn:parameters}. 
}
\label{fig:FTLM_lowTloc}
\end{figure}

\begin{figure*}
\includegraphics[width=.95\textwidth]{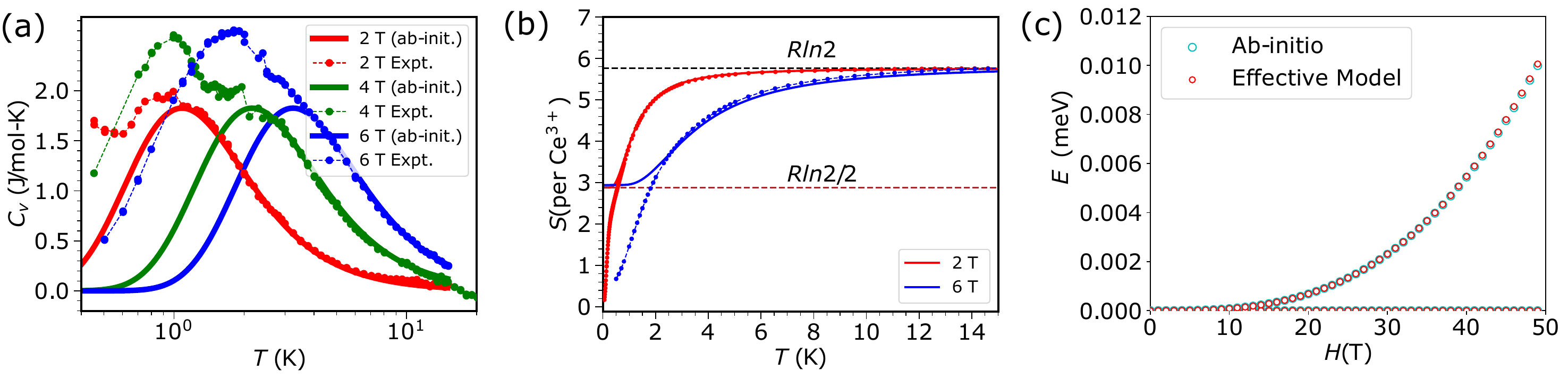}
\caption{
(a) Calculated specific heat using the \textit{ab-initio} Hamiltonian. (b) The entropy of a non-interacting Hamiltonian saturates at $R$ln 2/2 if the field is perfectly along the [110] direction. (c) Calculated first excited state using \textit{ab-initio} and effective theory (using $g_y\sim 0.0004$ meV$^{-2}$) for sublattice 2 in the presence of a [110] field.
\label{fig:cv_abinit}}       
\end{figure*}

\appendix
\section{Revisiting alternative scenarios}
\label{App:A}
\subsection{Details of the first scenario: Indirect coupling of $\beta$ spins}
\label{app:first_scenario}
We highlight the details of the framework structured by PMB who extensively studied the classical phase diagram of dipole-octupole pyrochlores in a [110] magnetic field for a wide range of parameter sets. Here we focus on the case where $J_y$ is dominant, since it is directly pertinent to the study of~\cezr~and~\cehf. It is worth noting that the argumentation would essentially remain similar if $J_x$ is dominant.

The gist of PMB's argument is as follows. For the sake of simplicity, consider the case of $g_x=0$ and turn off the transverse terms completely $(J_x=J_z=0)$, then each tetrahedron must satisfy the `ice rule' of $\sum_{i \epsilon \textrm{tet}} S^i_y = 0 $. Now if a large [110] magnetic field is applied, it polarizes the $\alpha$ spins - the sublattice 0 sites point in the local $+z$ direction and the sublattice 3 sites point in the local $-z$ direction, both have no local $y$ component. Sublattice 1 and 2 spins, which do not couple to the field, must then conspire to satisfy the ($y$-)ice rule. This can be done in two different ways: either 1 points in and 2 points out; or 1 points out and 2 points in, where ``in'' and ``out'' refer to local $y$ axes. Once this choice has been made for a single tetrahedron, it cascades onwards to cover the entire $\beta$ chain containing sublattice 1 and 2 spins. Since there are two such choices, one can assign an Ising degree of freedom to each $\beta$ chain. Each Ising spin is an independent degree of freedom, and thus the associated degeneracy is subextensive, scaling as the square of the linear system size as $L^2$ instead of its volume $N \sim L^3$ (where $N$ is the number of sites). 

The subextensive degeneracy is split by coupling between the $\beta$ chains, each representing an Ising spin, which themselves form an anisotropic triangular lattice. This coupling is a result of finite but small $J_x,J_z$ while keeping the [110] field large. In this regime, corresponding to the so-called `chain Y' phase, the calculated couplings between nearest neighbor Ising spins are given by Eq.~\ref{eq:PMB-pert}. As mentioned in section~\ref{sec:first_scenario}, this functional dependence suggests that the order-disorder transition of the $\beta$ chains should occur at \emph{lower} temperatures upon increasing the field strength, which is \emph{opposite} to the trend we observe in experiments.

We take this opportunity to clarify that the PMB procedure completely ignores quantum mechanical fluctuations within a $\beta$ chain. If these fluctuations are important, it is not apparent that it is justified to treat an entire $\beta$ chain as a single Ising degree of freedom. At large [110] fields the $\alpha$ spins are essentially locked and can be treated classically. However, sites on a single $\beta$ chain are interacting quantum mechanically with their immediate nearest neighbors on the same chain, and this intrinsically selects a state on the scale of the magnetic coupling strengths (which are at the scale of 0.4 K or less), essentially independent of the magnetic field strength. We see evidence for our assertion in Fig.~\ref{fig:FTLM_lowTloc} where we plot the location of the low temperature feature in the specific heat as a function of field strength. We see only a small \emph{increase} with increasing $h$ for low $h$, and essentially $h$ independence at large $h$. 

\subsection{Details of the second scenario: Higher order Zeeman coupling}
\label{app:second_scenario}
Here we discuss the details of the second scenario (mentioned in section~\ref{sec:second_scenario}) and the fitting procedure used to estimate $g_y$ in Eq.~\ref{eq:Hz}. As a first exercise we attempted to fit the effective Hamiltonian parameters, along with the additional $g_y$ term, to our specific heat data. We found $g_y$ to be somewhat large ($\approx 10$ meV$^{-2}$), and our fits could only reproduce the experimental features at low fields.

In order to assess the possible importance of higher-order field terms and extract the $g_y$ directly, we performed diagonalization of the single ion \textit{ab-initio} Hamiltonian which is a sum of the spin-orbit coupling term, the crystal field Hamiltonian and the Zeeman term in the space spanned by $J=5/2$ and $J=7/2$ manifold (i.e.,  Hilbert space of size 14). We ignore interactions between the ions. For a single ion, the form of the crystal field Hamiltonian is
\begin{eqnarray}
\label{eq:stevens}
H_\text{CF}&=& B_{20}O_{20}+B_{40}O_{40}+B_{43}O_{43}+B_{60}O_{60}+\nonumber \\
&&B_{63}O_{63}+B_{66}O_{66}.
\end{eqnarray}
For Ce$_2$Hf$_2$O$_7$, $B_{20}=-0.820(188)$, $B_{40}=0.223(5)$, $B_{43}=1.773(53)$, $B_{60}=-0.008(1)$, $B_{63}=0.074(2)$ and $B_{66}=-0.075(2)$ in meV. $O_{ij}$ are the Stevens operators~\cite{Stevens_Op,Wybourne}.  
These values of crystal field parameters were determined by fitting the CEF Hamiltonian parameters to inelastic neutron spectroscopy and magnetic susceptibility data measured on a powder sample of \cehf~as reported in \cite{cehfo-poree}. Although the fit was somewhat under-constrained, it showed good agreement with the experimental data and produced reasonable uncertainties on the CEF parameters, making them a suitable basis for our model. Furthermore, variations within these uncertainties are unlikely to significantly alter the wavefunction and the associated magnetic g-tensor.\\

The spin-orbit coupling part of the Hamiltonian is given by
\begin{eqnarray}
\label{eq:soc}
H_\text{SO}=\lambda_\text{SO} \;\; {\bf{L}}\cdot{\bf{S}}.
\end{eqnarray}
Here, $\lambda_\text{SO}$, which determines the energy scale of the spin-orbit interaction, was set at 80.3 meV. 

The bare Zeeman term is given by
\begin{eqnarray}
H_\text{Z}=-\mu_B\bf{h}\cdot(\bf{L}+2\bf{S}).
\end{eqnarray}
With the CEF parameters mentioned above, the specific heat for a single tetrahedron in the presence of a [110] field is plotted in Fig.~\ref{fig:cv_abinit}(a)  by using Eq.~\ref{eq:H}.
\begin{eqnarray}
H=H_\text{CF}+H_\text{SO}+H_\text{Z}.
\label{eq:H}
\end{eqnarray}
The \textit{ab-initio} Hamiltonian is able to capture the high-temperature part of the specific heat accurately.
It should be noted that at smaller values of field ($ \lesssim 10$ T) the Zeeman terms in the \textit{ab-initio} Hamiltonian for sublattice 1 and 2 ($\beta$ spins) are essentially zero and, therefore, the two sites do not contribute to the specific heat. Hence, the $\alpha$ spins alone explain the field-dependent behavior of the specific heat at higher temperatures. This is also clearly realized by noting that the evaluated entropy (per site) for the single tetrahedron as a function of temperature with the field along [110] direction and at very high temperature is $R\ln 2/2$ instead of $R \ln 2$ (see Fig.~\ref{fig:cv_abinit}b).

We confirm the finding of the weak coupling of the $\beta$ sites, by calculating $g_y$. To do so, we evaluate the energy splitting of the lowest lying doublet in the presence of [110] field using Eq.~\ref{eq:H}. By fitting to an appropriately defined cost function, this energy difference curve directly gives us $g_y$, which we find to be $0.0004$ meV$^{-2}$. 

In Fig.~\ref{fig:cv_abinit}c we show the energy splitting for sublattice 2 obtained by using the \textit{ab-initio} Hamiltonian and from the effective theory with $g_y\sim 0.0004$ meV$^{-2}$. Even at 50 Tesla the energy splitting of the doublet is only of the order of 0.01 meV, suggesting that the system is still in the perturbative regime, and that the estimate of $g_y\sim 10$ meV$^{-2}$ from directly fitting the low field specific heat data is unrealistic.

\begin{figure}
\includegraphics[width=0.45\textwidth, angle=-90]{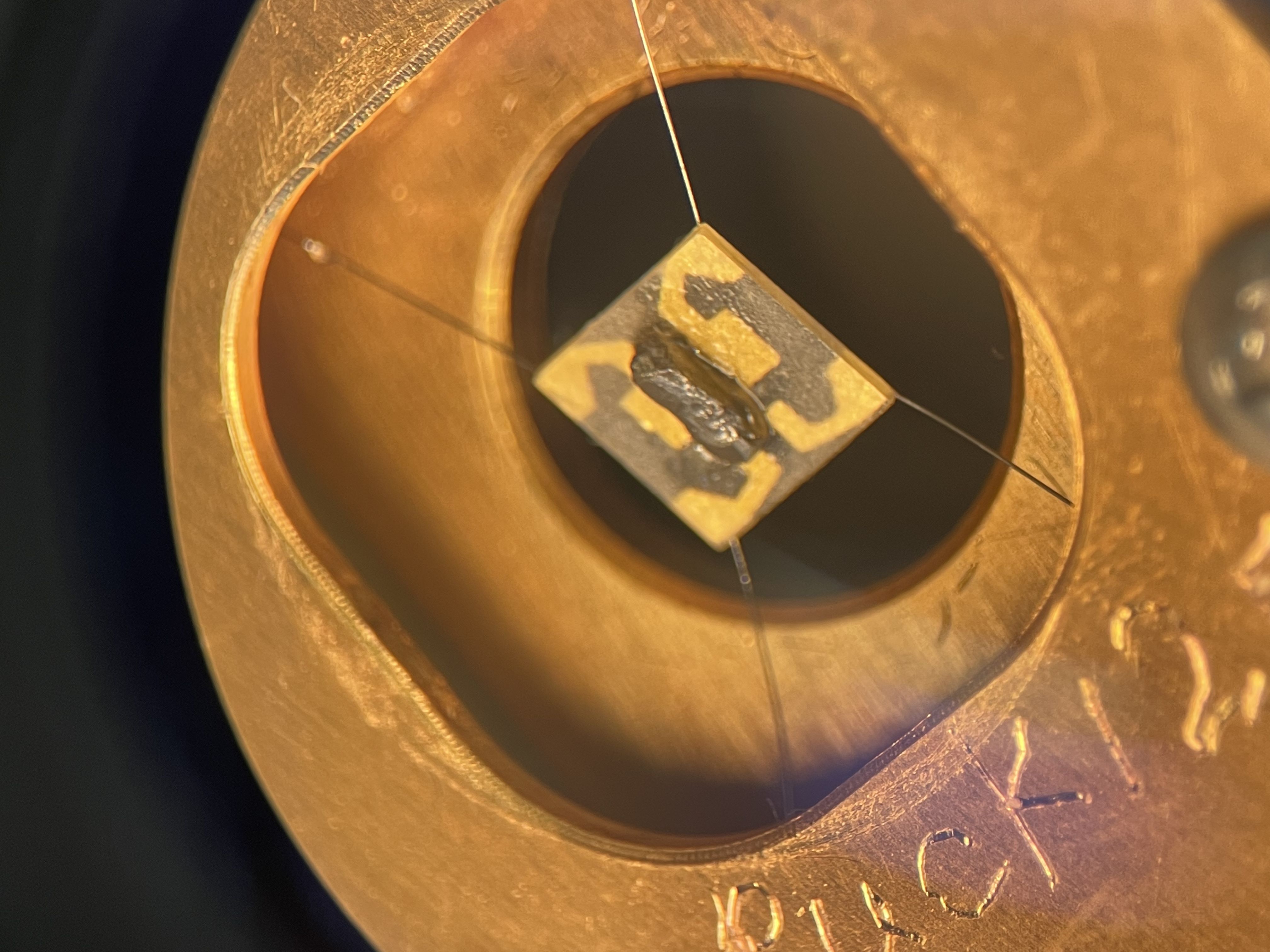}
\caption{
Photograph of a HoGa sample mounted on the PPMS $^{3}$He specific heat platform, after measurements at low temperature under moderate magnetic field. The initial orientation of the sample platform was in the same plane as copper surface on the puck. We note that the specific heat puck remained fully functional, both during and after the tilt triggered by the magnetic torque of the sample.}
\label{fig:PPMSplatform}  
\end{figure}

\begin{figure*}
\includegraphics[width=0.95\textwidth]{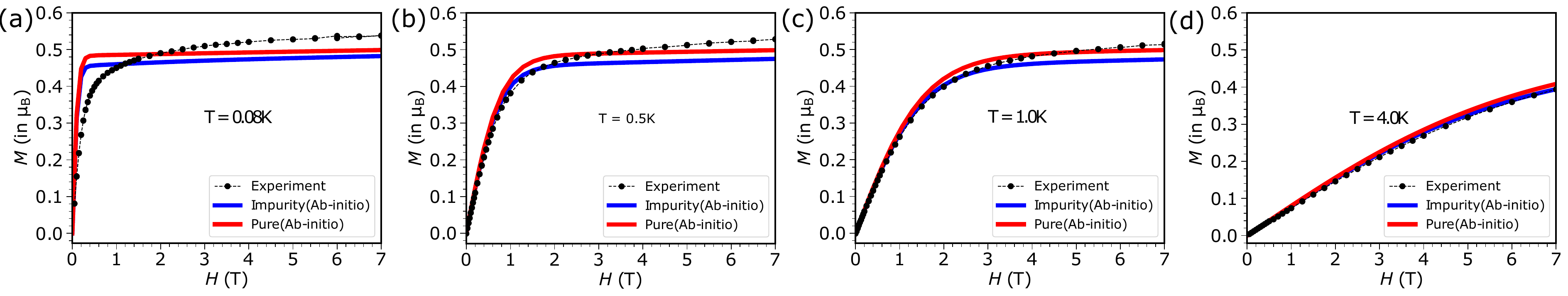}
\caption{Magnetization from \textit{ab-initio} (PyCEF) calculations with and without impurities at four representative temperatures. The black solid spheres connected by dashed lines represent the experimental data, the red and blue solid lines represent the magnetization obtained from PyCEF in the absence and presence of 10\% nonmagnetic Ce$^{4+}$ respectively. The impurity (blue) curve was obtained by scaling contributions from two kinds of Ce$^{3+}$ (30\% in a pure environment and 60\% in a defective environment), as explained in the text.}
\label{fig:appfig2}   
\end{figure*}

\section{Specific heat measurements at high fields}
\label{App:B}
The specific heat data presented in this work, for the highest magnetic fields, were recorded using the PPMS $^{3}$He specific heat setup. From our analysis, we argue that relatively large tilts of the sample platform can be induced by the torque of the sample when applying magnetic field in cryogenic conditions. As an example of this we show in Fig.~\ref{fig:PPMSplatform}  the PPMS $^{3}$He specific heat platform after an unrelated experiment performed using a similar setup on a sample of composition HoGa. The sample was cooled down without magnetic field, after which a moderate magnetic field was applied for measurements, then removed before warming up the sample. The tilt of the sample platform is approximately 30 degrees.  In such an extreme case, the sample torque triggers changes to the platform going beyond the `plastic limit' of the assembly, which thus can be observed after taking the setup out of the cryomagnet. The value of $\phi_\text{opt}$ deduced from the present experiments on \cehf~remain smaller than what is exemplified in Fig.~\ref{fig:PPMSplatform}.

\section{Impact of nonmagnetic impurities on magnetization}
\label{App:C}
In order to capture the behavior of the magnetization as a function of a magnetic field applied along the [110] direction, we also considered the effect of the non-magnetic Ce$^{4+}$ impurities. To do so, we used the two CEF models developed in \cite{cehfo-poree}: one where the Ce$^{3+}$ ion experiences an ideal D$_{3d}$ environment and a second one where the D$_{3d}$ symmetry is broken. These two models were subsequently fitted to the INS and magnetic susceptibility data \cite{cehfo-poree}. A notable difference between the two models is the reduced Ising anisotropy of the model with impurities, which translates into a $g$-tensor containing more non-zero contributions beyond $g_{zz}$. 

A system with $N$ sites has 0.25$N$ number of spins on each sublattice. 
We assume a dilute density of impurities (in the form of nonmagnetic Ce$^{4+}$ in the ballpark of 10$\%$) that are homogeneously distributed - this means 0.025$N$ sites on each sublattice are nonmagnetic. 
If these impurities are far from each other and any magnetic site at most sees only a single impurity, then 0.15$N$ (0.075$N$) sites of each sublattice type experience a ``defective" (perfect) crystal field environment; this is because any single impurity on any sublattice will affect the environment of 6 neighboring cerium atoms. This means that 60\% of Ce experience a defective environment, and the remaining 30\% experience a perfect CEF environment.

If this effect were dominant, we anticipate that summing the magnetization curves scaled by 0.3 and 0.6 for the pure and impurity models, respectively, would reproduce the experimental data. However, we find that this approach falls short of capturing the measured magnetization, as can be seen in Fig.\ref{fig:appfig2}. As expected, the high-temperature data are reasonably well captured by this model, since exchange interactions are negligible in this regime. However, the low-temperature data are \textit{not} well reproduced, particularly at low field strengths where the exchange terms are dominant. Additionally, at intermediate temperatures, not only is the magnitude of the computed high-field magnetization too small, but the computed slope (above 4~T) is also smaller than that observed in the experiment. 

Based on these observations, we argue that although the presence of impurities does affect the magnetization, its impact is minor and alone does not accurately reproduce our experimental data.

\bibliography{refs}

\end{document}